\documentclass[12pt]{article}

\usepackage[square,comma,numbers,sort&compress]{natbib}
\usepackage[left=20mm,right=20mm,top=20mm,bottom=20mm]{geometry}
\usepackage[dvipdfmx]{graphicx}
\usepackage{mediabb}

\usepackage{epsf}
\usepackage{color}
\usepackage{amsmath}
\usepackage{amssymb}
\usepackage{latexsym}
\usepackage{slashed}
\usepackage{float}
\usepackage{cases}
\usepackage{multirow}
\usepackage{bm}
\usepackage{braket}

\newcommand{\al}[1]{\begin{align}#1\end{align}}

\newcommand{\bp}{\begin{pmatrix}}
\newcommand{\ep}{\end{pmatrix}}
\newcommand{\bb}{\begin{bmatrix}}
\newcommand{\eb}{\end{bmatrix}}

\newcommand{\fulltoday}{\number\day\space \ifcase\month\or
    January\or February\or March\or April\or May\or June\or
    July\or August\or September\or October\or November\or December\fi
    \space\number\year}

 \makeatletter
    \renewcommand{\theequation}{%
    \thesection.\arabic{equation}}
    \@addtoreset{equation}{section}
  \makeatother

\begin{document}
\allowdisplaybreaks[2]
\begin{titlepage}
\renewcommand\thefootnote{\alph{footnote}}
		\mbox{}\hfill EPHOH-16-004\\
		\mbox{}\hfill KIAS-P16032\\
		\mbox{}\hfill KOBE-TH-16-03\\
		\mbox{}\hfill WU-HEP-16-07\\
\vspace{2mm}
\begin{center}
{\fontsize{22pt}{0pt}\selectfont \bf{Comprehensive Analysis of Yukawa Hierarchies on $T^2/Z_N$ with Magnetic Fluxes}
} \\
\vspace{8mm}
	{}
	{\fontsize{14pt}{0pt}\selectfont \bf
	Yukihiro Fujimoto,\,\footnote{
		E-mail: \tt fujimoto@oita-ct.ac.jp
		}}
	{}
	{\fontsize{14pt}{0pt}\selectfont \bf
	Tatsuo Kobayashi,\,\footnote{
		E-mail: \tt kobayashi@particle.sci.hokudai.ac.jp
		}}
	{}
	{\fontsize{14pt}{0pt}\selectfont \bf
	Kenji Nishiwaki,\,\footnote{
		E-mail: \tt nishiken@kias.re.kr
		}}
	{}
	\\[3pt]
	{\fontsize{14pt}{0pt}\selectfont \bf
	Makoto Sakamoto,\,\,\footnote{
		E-mail: \tt dragon@kobe-u.ac.jp} }
	{\fontsize{14pt}{0pt}\selectfont \bf
	Yoshiyuki Tatsuta\,\,\footnote{
		E-mail: \tt y{\_}tatsuta@akane.waseda.jp} } \\
\vspace{4mm}
	{\fontsize{13pt}{0pt}\selectfont
		${}^{\mathrm{a}}$\it National Institute of Technology, Oita College, Oita 870-0152, Japan \smallskip\\[5pt]
		${}^{\mathrm{b}}$\it Department of Physics, Hokkaido University, Sapporo 060-0810, Japan \smallskip\\[5pt]
		${}^{\mathrm{c}}$\it School of Physics, Korea Institute for Advanced Study,\\[3pt]
					85 Hoegiro, Dongdaemun-gu, Seoul 02455, Republic of Korea\\[5pt]
		${}^{\mathrm{d}}$\it Department of Physics, Kobe University, Kobe 657-8501, Japan \smallskip\\[5pt]
		${}^{\mathrm{e}}$\it Department of Physics, Waseda University, Tokyo 169-8555, Japan \smallskip\\[5pt]
	}
\vspace{4mm}
{\normalsize \fulltoday}
\vspace{10mm}
\end{center}
\begin{abstract}
\fontsize{12pt}{16pt}\selectfont{
{Based on the result of classification in Ref.~\cite{Abe:2015yva}, we exhaustively investigate Yukawa sector of $U(8)$ model on magnetized orbifolds $T^{2}/Z_{2}$, $T^{2}/Z_{3}$, $T^{2}/Z_{4}$ and $T^{2}/Z_{6}$ by evaluating ratios of the mass eigenvalues of the three states in all the possible configurations with one and two Higgs pairs where three generations are realized in fermions.
Because of smearing effect via kinetic mixing, {one can realize a hierarchy such as $10^{-2}$-$10^{-3}$, but}
it is very difficult to achieve the mass ratio between the up and top quarks ($m_{\text{up}}/m_{\text{top}} \sim 10^{-5}$)} on the complicated magnetized orbifolds $T^{2}/Z_{N}\,(N=3,4,6)$.
}
\end{abstract}
\end{titlepage}
\renewcommand\thefootnote{\arabic{footnote}}
\setcounter{footnote}{0}


\section{Introduction}

Even after the completion of the standard model~(SM) by the discovery of the Higgs boson~\cite{Aad:2012tfa,Chatrchyan:2012ufa}, the origin of properties of the matter fields, especially the very hierarchical mass spectra of the quarks and the leptons is still concealed.
Quantized magnetic fluxes of a {\it unified} gauge group among {extra} directions of the spacetime provide us a fascinating guiding principle, where the fluxes trigger an explicit breaking of a unified gauge group down to the ones including the SM gauge group.
Interestingly, the {four-dimensional} chiral fermions on such magnetized backgrounds are degenerated and their profiles are quasi-localized in the extra dimensions, where the former and latter properties give us reasonable answers to the mysteries of the three generations and the Yukawa hierarchies, respectively.
A reasonable starting point is ten-dimensional (10D) super Yang-Mills~(SYM) theory on magnetized tori~\cite{Bachas:1995ik,Blumenhagen:2000wh,Angelantonj:2000hi,Blumenhagen:2000ea,Cremades:2004wa,Blumenhagen:2005mu,Blumenhagen:2006ci}, which could be an effective theory of superstring theories.\footnote{In this paper, our setups will be discussed in a framework of supersymmetric Yang-Mills theories with $U(N)$ gauge groups. It would, however, be of great interest to embed our setups into superstring theories. Although it is known that $U(N)$ gauge groups are hard to be obtained in 10D type II string theory with D-branes, our results will be still useful because some properties with respect to zero-mode {wave functions}, their degeneracies and Yukawa couplings induced by magnetic fluxes are the same as those even in cases of other gauge groups.}
Note that various phenomenological aspects have been pursued: Yukawa couplings~\cite{Cremades:2004wa},
realization of quark/lepton masses and their mixing angles~\cite{Abe:2012fj,Abe:2014vza}, higher order couplings~\cite{Abe:2009dr}, flavor symmetries~{\cite{Abe:2009vi,Abe:2009uz,Abe:2010ii,BerasaluceGonzalez:2012vb,Honecker:2013hda,Marchesano:2013ega,Abe:2014nla}},
massive modes~\cite{Hamada:2012wj}, and others~\cite{Sakamoto:2003rh,Antoniadis:2004pp,Antoniadis:2009bg,Choi:2009pv,Kobayashi:2010an,DiVecchia:2011mf,Abe:2012ya,DeAngelis:2012jc,Abe:2013bba,Hamada:2014hpa,Abe:2015jqa,Sumita:2015tba,Abe:2015bxa,Kobayashi:2015siy}.
Some related works in other stringy contexts ({e.g.,} intersecting D-brane model and heterotic string theory) are found in~\cite{Cremades:2003qj,Cvetic:2003ch,Abel:2003vv,Kobayashi:2004ya,Blumenhagen:2005mu,Blumenhagen:2006ci,Kobayashi:2006wq,Ko:2007dz,Honecker:2012jd,Beye:2014nxa,Abe:2015mua,Abe:2015xua,Abe:2015uma}.\footnote{
Another attractive direction is considering various boundary conditions of fields on point interactions (zero-thickness branes) in the bulk space of a five-dimensional theory on $S^1$ (or a line segment)~\cite{Fujimoto:2012wv,Fujimoto:2013ki,Fujimoto:2014fka,Fujimoto:2016gfu}.
}
Two-dimensional torus is the simplest choice in the space which the magnetic fluxes are turned on.
Besides, various toroidal orbifolds can be candidates.
Note that their geometrical aspects are discussed within the context of string theory~\cite{Katsuki:1989bf,Kobayashi:1991rp,Choi:2006qh} and higher-dimensional field theories~\cite{Kawamura:2008mz,Kawamura:2007cm,Kawamura:2009sa,Kawamura:2009gr,Goto:2013jma,Goto:2014eoa,Goto:2015iaa}.

Previously, a technical difficulty exists for analytical calculations of Yukawa couplings when the backgrounds are both magnetized and orbifolded.
However, a progress was made in Ref.~\cite{Abe:2014noa} in a {\it dual} description with operator formalism,\footnote{
Note that only $T^{2}$-related cases were discussed in~\cite{Abe:2014noa}.
But, addressing other higher-dimensional tori is possible in principle in a similar method.
The case of shifted orbifolds with magnetic fluxes was done in~\cite{Fujimoto:2013xha}.
}
which enables us to write down analytical forms of Yukawa couplings on such complicated geometries even though Wilson line phases and/or Scherk-Schwarz phases are also introduced (see~\cite{Abe:2013bca} and~\cite{Scherk:1978ta,Scherk:1979zr,Kobayashi:1991rp,Ibanez:1986tp,Kobayashi:1990mi,Angelantonj:2005hs,Blumenhagen:2005tn,Angelantonj:2009yj,Forste:2010gw,Nibbelink:2012de}.).
Inspired by this achievement, all of the possible configurations were derived and classified when the gauge group is $U(8)$ in Ref.~\cite{Abe:2015yva}, which is the minimal choice in the case of $U(N)$-type theories~\cite{Abe:2008fi,Abe:2008sx,Abe:2012fj,Abe:2013bba,Abe:2014vza} {(see also~\cite{Choi:2009pv,Kobayashi:2010an} and~\cite{Abe:2015mua,Abe:2015xua} for discussions in $E_{6,7,8}$ and $SO(32)$ groups, respectively.)}.

Based on the result of classification, we exhaustively analyze all the possible configurations with one and two Higgs pairs on orbifolds $T^{2}/Z_{2},\,T^{2}/Z_{3},\,T^{2}/Z_{4},\,T^{2}/Z_{6}$ where three generations are realized in fermions by evaluating ratios of the mass eigenvalues of the three states.\footnote{
A similar calculation was done in the context of gauge-Higgs unification scenario in the gauge group $G \times U(1)_{X}$ on $T^{2}/Z_{N}\,(N=2,3,4,6)$, where $G$ is a simple group including the part $SU(2)_{L} \times U(1)_{Z}$ in~\cite{Matsumoto:2016okl} by adopting the same method in~\cite{Abe:2014noa,Abe:2015yva}.
Note that matter unification is not realized in this model.
}
An important point is that we should realize the ratio $m_{\text{up}}/m_{{\text{top}}} \sim 10^{-5}$ in the up-quark sector.
Whether or not this magnitude is achievable is a significant criterion for selecting type of configurations on the backgrounds, $T^{2}/Z_{2},\,T^{2}/Z_{3},\,T^{2}/Z_{4},\,T^{2}/Z_{6}$.

This paper is organized as follows.
In Sec.~\ref{sec:review}, we briefly review basic properties of magnetized systems, which {include} the explicit form for Yukawa calculations on magnetized orbifolds based on two-dimensional torus.
In Sec.~\ref{sec:result}, we calculate ratios of mass eigenvalues of fermion zero modes in all the configurations with three generations and one or two pairs of $SU(2)_{L}$ Higgs doublets.
{Section~\ref{sec:conclusion} is the} conclusion.

\section{Brief review on 10D SYM on magnetized orbifolds
\label{sec:review}}

\subsection{Setups}

First of all, we briefly review the basics of the {ten-dimensional} super Yang-Mills theory on (generalized) magnetized orbifolds {based on descriptions in~\cite{Abe:2015yva}}, where we focus on the {two-dimensional} part determining flavor structure of ${T^{2}/Z_{N}}\, (N=2,3,4, \text{or} \ 6)$.
We consider the $U(N)$ theory in the notation adopted in~\cite{Abe:2015yva},
\al{
S = \int_{M^4} d^4x \int_{(T^2)^3} d^6z \left\{ -\frac{1}{4} \text{tr} \left( F_{MN}F^{MN} \right) +
	\frac{1}{2} \text{tr} \left( \overline{\lambda} \Gamma^M i D_M \lambda \right) \right\},
	\label{10D_action}
}
which is defined on a product of {four-dimensional} Minkowski space and three factorizable {$2$-tori}.
The capital roman indices $M,N$ run over $\mu \,(=0,1,2,3)$, $\{z_i, \overline{z_i}\}$, where the $i$-th ($i=1,2,3$) 2-torus is described by the complex {coordinates} $z_i = y_{2i+2} + i y_{2i+3}$ and its complex conjugation $\overline{z_i} = y_{2i+2} - i y_{2i+3}$ made by the two Cartesian {coordinates} representing the extra directions, $y_{2i+2}$ and $y_{2i+3}$.
We take each torus modulus parameter $\tau_i\ (\subset \mathbb{C})$ as $\text{Im}\tau_i > 0$ for convenience. 
We use the short-hand notation $d^6 z$ meaning $\Pi_{i=1}^3 dz_i d\overline{z_i}$.
On $T^2_i$, the coordinate $z_i$ is identified as $z_i \sim z_i+1 \sim z_i+\tau_i$.
The bulk $\mathcal{N}=1$ supersymmetric (in {ten-dimensional}) theory contains the {ten-dimensional} vector fields $A_{M}$ and the gaugino fields $\lambda$ described by {ten-dimensional} Majorana-Weyl spinors.

The gaugino fields and the {ten-dimensional} vector fields are Kaluza-Klein (KK) decomposed as
\al{
\lambda(x, \{z_i, \overline{z_i}\}) &=
	\sum_{l,m,n} \chi_{l,m,n}(x) \otimes \psi^{(1)}_l (z_1, \overline{z_1}) \otimes
	\psi^{(2)}_m (z_2, \overline{z_2}) \otimes \psi^{(3)}_n (z_3, \overline{z_3}), \\
A_M(x, \{z_i, \overline{z_i}\}) &=
	\sum_{l,m,n} \varphi_{l,m,n; M}(x) \otimes \phi^{(1)}_{l,M} (z_1, \overline{z_1}) \otimes
	\phi^{(2)}_{m,M} (z_2, \overline{z_2}) \otimes \phi^{(3)}_{n,M} (z_3, \overline{z_3}),
}
where $l,m,n$ are KK indices and $\psi^{(i)}_l$ is a {two-dimensional} spinor describing the {$l$th} KK mode on the {$i$th} $T^2$, whose exact form is $\psi^{(i)}_l = \left( \psi^{(i)}_{l,+}, \psi^{(i)}_{l,-} \right)^\text{T}$ and the corresponding {two-dimensional} chirality ($+$ or $-$) is denoted by $s_i$.
We {adopt} the gamma matrices $\tilde{\Gamma}^m$ (identified by the Cartesian coordinates) corresponding to the {$i$th} torus as
\al{
\tilde{\Gamma}^{2i+2} = i\sigma_1,\quad
\tilde{\Gamma}^{2i+3} = i\sigma_2,
}
where {$\sigma_{1,2}$ are Pauli matrices}.
In the following part, we basically focus on the zero modes $(l=m=n=0)$ on the $T^{2}/Z_{N}$ sector {which} describes flavor structure, by omitting the KK and torus indices.
We also {skip} to show coordinates $z_{2}, \bar{z}_{2}, z_{3}, \bar{z}_{3}$ in {ten-dimensional} fields to avoid clumsy descriptions.

We introduce factorizable Abelian magnetic fluxes on the three $T^2$ through the classical vector potential of $A_M$ in the following forms
\al{
A^{(b)}(z, \overline{z}) &= \frac{\pi}{{g} \text{Im} \tau}
\begin{pmatrix}
M_1 \text{Im}\left[ (\overline{z} + \overline{C_1}) dz \right] \mathbf{1}_{N_1 \times N_1} & & 0 \\
& \ddots & \\
0 & & M_{n} \text{Im}\left[ (\overline{z} + \overline{C_n}) dz \right] \mathbf{1}_{N_n \times N_n}
\end{pmatrix} \notag \\
	&= \frac{1}{4 i \text{Im} \tau} \left( \overline{{\boldsymbol B}} dz - {\boldsymbol B} d\overline{z} \right), 
	\label{1-form_potential_general}
}
where $C_j\ (j=1,\cdots,n)$ represent the corresponding Wilson line phases on $T^2/Z_{N}$, and $M_j\ (j=1,\cdots,n)$ should be integers because of Dirac's quantization condition on $T^2/Z_{N}$.
On the magnetized $T^2/Z_{N}$, possible choices of $C_j$ are limited.
Since it was shown that this degrees of freedom can be gauged away by a large gauge transformations~\cite{Abe:2013bca}, we set $C_{j} = 0$.
Under this background, the original gauge group $U(N)$ explicitly breaks down as $U(N) \to \Pi_{a=1}^n U(N_a)$ with $N = \sum_{a=1}^n N_a$.
We can derive the following relations,
\al{
A^{(b)}{( z + 1, \overline{z} + 1 )} &= A^{(b)}(z, \overline{z}) + {d} \xi_{1}(z), \\
A^{(b)}{( z + \tau, \overline{z} + \overline{\tau} )} &= A^{(b)}(z, \overline{z}) + {d} \xi_{\tau}(z), \\
{\xi_{1}(z)} &= \frac{1}{2\text{Im}\tau} \text{Im}[{\boldsymbol B}],\\
{\xi_{\tau}(z)} &= \frac{1}{2\text{Im}\tau} \text{Im}[\overline{\tau}{\boldsymbol B}],
}
where {$d$} plays as an exterior derivative on {$T^2/Z_{N}$}.
Here, the Lagrangian density in Eq.~(\ref{10D_action}) should be single-valued under every torus identification $z \sim z +1 \sim z + \tau$, and then in the gaugino fields $\lambda(x, z, \overline{z})$, the following pseudo-periodic boundary conditions should be arranged,
\al{
\lambda{(x, z + 1, \overline{z} + 1)} &= {U_{1}(z)} \lambda(x, z, \overline{z}) {U_{1}(z)^\dagger}, \\
\lambda{(x, z + \tau, \overline{z} + \overline{\tau})} &= {U_{\tau}(z)} \lambda(x, z, \overline{z}) {U_{\tau}(z)^\dagger},
}
with
\al{
U_{1}(z) := e^{i {g} \xi_{1}(z) + 2\pi i {\boldsymbol \alpha}}, \quad
U_{\tau}(z) := e^{i {g} \xi_{\tau}(z) + 2\pi i {\boldsymbol \beta}},
	\label{boundarycondition_general}
}
\vspace{-6mm}
\al{
{\boldsymbol \alpha} :=
\begin{pmatrix}
\alpha_{1} \mathbf{1}_{N_1 \times N_1} & & 0 \\
& \ddots & \\
0 & & \alpha_{n}  \mathbf{1}_{N_n \times N_n}
\end{pmatrix},\quad
{\boldsymbol \beta} :=
\begin{pmatrix}
\beta_{1} \mathbf{1}_{N_1 \times N_1} & & 0 \\
& \ddots & \\
0 & & \beta_{n}  \mathbf{1}_{N_n \times N_n}
\end{pmatrix},
	\label{SS_matrix}
}
where $\alpha_{j}$ and $\beta_{j}$ $(j=1,\cdots,n)$ describe Scherk-Schwarz phases and can take limited numbers.

Here, we exemplify the fermionic part in the specific case of $U(N) \to U(N_a) \times U(N_b)$.
As {two-dimensional} spinors, the gaugino fields are decomposed as
\al{
\psi(z, \overline{z}) =
\begin{pmatrix}
\psi^{aa}(z, \overline{z}) & \psi^{ab}(z, \overline{z}) \\
\psi^{ba}(z, \overline{z}) & \psi^{bb}(z, \overline{z})
\end{pmatrix}.
}
The parts represented by $\psi^{aa}$ and $\psi^{bb}$ correspond to the representation under the unbroken gauge group $U(N_a) \times U(N_b)$, while $\psi^{ab}$ and $\psi^{ba}$ are the {bifundamental} matter fields as $(N_a, \overline{N_b})$ and $(\overline{N_a}, N_b)$, respectively.
We obtain the zero-mode equations for these gaugino fields on the $T^2/Z_{N}$ with the {two-dimensional} chirality ($+$ or $-$) as
\al{
\begin{pmatrix}
\partial_{\overline{z}} \psi^{aa}_{+} &
\left[ \partial_{\overline{z}} + \frac{\pi}{2\text{Im} \tau} \left( M_{ab} z \right) \right] \psi^{ab}_{+} \\
\left[ \partial_{\overline{z}} + \frac{\pi}{2\text{Im} \tau} \left( M_{ba} z \right) \right]  \psi^{ba}_{+} &
\partial_{\overline{z}} \psi^{bb}_{+}
\end{pmatrix} &= 0,
\label{2D_gauginoequation_plus} \\
\begin{pmatrix}
\partial_{z} \psi^{aa}_{-} &
\left[ \partial_{z} - \frac{\pi}{2\text{Im} \tau} \left( M_{ab} \overline{z} \right) \right] \psi^{ab}_{-} \\
\left[ \partial_{z} - \frac{\pi}{2\text{Im} \tau} \left( M_{ba} \overline{z} \right) \right]  \psi^{ba}_{-} &
\partial_{z_i} \psi^{bb}_{-}
\end{pmatrix} &= 0,
\label{2D_gauginoequation_minus}
}
with the short-hand notations $M_{ab} := M_{a} - M_{b}$.
The effective boundary conditions of the fields are easily written down,
\al{
\psi^{ab}_{s}{(z +1, \overline{z} +1)} &= e^{i\frac{\pi s}{\text{Im} \tau} \text{Im} \left[ M_{ab} z \right] + 2 \pi i \alpha_{ab}} \psi^{ab}_{s}{(z, \overline{z})}, \notag \\
\psi^{ba}_{s}{(z +1, \overline{z} +1)} &= e^{i\frac{\pi s}{\text{Im} \tau} \text{Im} \left[ M_{ba} z \right] + 2 \pi i \alpha_{ba}} \psi^{ba}_{s}{(z, \overline{z})}, \notag \\
\psi^{aa}_{s}{(z +1, \overline{z} +1)} &= \psi^{aa}_{s}{(z, \overline{z})}, \notag \\
\psi^{bb}_{s}{(z +1, \overline{z} +1)} &= \psi^{bb}_{s}{(z, \overline{z})},\\[4pt]
\psi^{ab}_{s}{(z + \tau, \overline{z} + \overline{\tau})} &= e^{i\frac{\pi s}{\text{Im} \tau} \text{Im} \left[ \overline{\tau} \left( M_{ab} z \right) \right] + 2 \pi i \beta_{ab}} \psi^{ab}_{s}{(z, \overline{z})}, \notag \\
\psi^{ba}_{s}{(z + \tau, \overline{z} + \overline{\tau})} &= e^{i\frac{\pi s}{\text{Im} \tau} \text{Im} \left[ \overline{\tau} \left( M_{ba} z \right) \right] + 2 \pi i \beta_{ba}} \psi^{ba}_{s}{(z, \overline{z})}, \notag \\
\psi^{aa}_{s}{(z + \tau, \overline{z} + \overline{\tau})} &= \psi^{aa}_{s}{(z, \overline{z})}, \notag \\
\psi^{bb}_{s}{(z + \tau, \overline{z} + \overline{\tau})} &= \psi^{bb}_{s}{(z, \overline{z})},
}
with the short-hand notations $\alpha_{ab} := \alpha_{a} - \alpha_{b}$ and $\beta_{ab} := \beta_{a} - \beta_{b}$.
We {note} that $s$ shows the corresponding {two-dimensional} chirality.

On $T^2$, possible twisted orbifolding is to impose the covariance on the fields under the rotation with the angle $\omega$, $z \to \omega z$, where $\omega$ is $e^{2\pi i/N}$ with $N=2,3,4,6$.
In other words, $Z_2$, $Z_3$, $Z_4$ and $Z_6$ (twisted) orbifoldings are realizable on $T^2$.
In non-Abelian gauge theories, a nontrivial gauge structure part $P$ appears in the $Z_N$ manipulation as
\al{
A_{\mu}{(x, \omega z, \overline{\omega} \overline{z} )} &= P A_{\mu}(x, z, \overline{z}) P^{-1}, \\
A_{z}{(x, \omega z, \overline{\omega} \overline{z})} &= \overline{\omega} P A_{z}(x, z, \overline{z}) P^{-1}, \\
A_{\overline{z}}{(x, \omega z, \overline{\omega} \overline{z})} &= \omega P A_{\overline{z}}(x, z, \overline{z}) P^{-1}, \\
\lambda_{s=+}{(x, \omega z, \overline{\omega} \overline{z} )} &= P \lambda_{s=+}(x, z, \overline{z}) P^{-1}, \\
\lambda_{s=-}{(x, \omega z, \overline{\omega} \overline{z} )} &= \omega P \lambda_{s=-}(x, z, \overline{z}) P^{-1} \label{ZNcondition_lambda_minus},
}
where $P$ should satisfy the conditions, $P \in U(N)$ and  $P^N = {\mathbf 1_{N \times N}}$.
Here, to prevent an additional explicit gauge symmetry breaking via the orbifoldings, we should take the following form in $P$,
\al{
P = \begin{pmatrix}
\eta_1 {\mathbf 1_{N_1 \times N_1}} & & 0 \\
& \ddots & \\
0 & & \eta_n {\mathbf 1_{N_n \times N_n}}
\end{pmatrix},
}
with $\eta_j = \{1,\omega,\cdots,\omega^{N-1}\} \ (j \,{\in}\, 1,\cdots,n)$.
Within the concrete example of $U(N) \to U(N_a) \times U(N_b)$ discussed in the previous subsection, $\psi^{aa}_{+}$ and $\psi^{bb}_{+}$ have trivial $Z_N$ parity {irrespective of} the values of $\eta_a$ and $\eta_b$, while $\psi^{ab}_{+}$ and $\psi^{ba}_{+}$ can contain nontrivial values of $\eta_a \overline{\eta_b}$, $\overline{\eta_a} \eta_b$, respectively.
The conditions for the {two-dimensional} gauginos with negative chirality are evaluated with ease by use of the relation in Eq.~(\ref{ZNcondition_lambda_minus}).

\subsection{Yukawa coupling on magnetized $T^{2}$}

Before we go for issues on magnetized $T^{2}$ with orbifolding, we summarize how to calculate Yukawa couplings in which orbifolding is not imposed.

When $M_{ab} > 0$, the fields $\psi^{ab}_{+}$ and $\psi^{ba}_{-}$ contain $|M_{ab}|$ normalizable zero modes, while the others $\psi^{ba}_{+}$ and $\psi^{ab}_{-}$ have no corresponding one.
On the other hand in $M_{ab} < 0$, $|M_{ab}|$ normalizable zero modes are generated from each of $\psi^{ba}_{+}$ and $\psi^{ab}_{-}$, whereas there is nothing {from} $\psi^{ab}_{+}$ and $\psi^{ba}_{-}$.
In the case of $M_{ab} = 0$, like $\psi^{aa}_{s}$ or $\psi^{bb}_{s}$, only one {nonlocalized} mode is generated from each of the all sectors and {nothing of phenomenological interests occurs}.
When $M_{ab} > 0$, which is equal to $M_{ba} < 0$, the {wave functions} of $\psi^{ab}_{+}$ and $\psi^{ba}_{-}$ take the following forms:
\al{
\psi^{ab}_{+}(z) =
\sum_{I=0}^{|M_{ab}|-1}
\begin{pmatrix} \Theta^{(I + \alpha_{ab}, \beta_{ab})}_{M_{ab}} (z, \tau) \\ 0 \end{pmatrix}, \quad
\psi^{ba}_{-}(\overline{z}) =
\sum_{I=0}^{|M_{ba}|-1}
\begin{pmatrix} 0 \\ \Theta^{(I + \alpha_{ba}, \beta_{ba})}_{M_{ba}} (\overline{z}, \overline{\tau})\end{pmatrix},
}
\al{
\Theta^{(I + \alpha_{ab}, \beta_{ab})}_{M_{ab}} (z, \tau) &=
\mathcal{N}_{|M_{ab}|}
e^{i\pi M_{ab} \, {z} \, {\mathrm{Im}(z)\over \mathrm{Im}\tau}}\cdot \vartheta \left[
\begin{array}{c}
{I + \alpha_{ab} \over M_{ab}} \\ -\beta_{ab}
\end{array}
\right] (M_{ab} \, {z}, M_{ab} \, \tau), \\
\Theta^{(I + \alpha_{ba}, \beta_{ba})}_{M_{ba}} (\overline{z}, \overline{\tau}) &=
\mathcal{N}_{|M_{ba}|}
e^{i\pi M_{ba} \, {\overline{z}} \, {\mathrm{Im}(\overline{z})\over \mathrm{Im} \overline{\tau}}}\cdot \vartheta \left[
\begin{array}{c}
{I + \alpha_{ba} \over M_{ba}} \\ -\beta_{ba}
\end{array}
\right] (M_{ba} \, {\overline{z}}, M_{ba} \, \overline{\tau}).
}
Here, $I\ (=0,\cdots,|M_{ab}|-1)$ discriminates the $|M_{ab}|$-degenerated zero-mode states.
The ({generalized}) $\vartheta$ function is defined by 
\begin{align}
&\vartheta \left[
\begin{array}{c}
a\\ b
\end{array}
\right] (c\nu,c\tau) 
=\sum_{l=-\infty}^{\infty}e^{i\pi (a+l)^2c\tau}e^{2\pi i(a+l)(c\nu +b)} ,
\end{align}
with the properties
\begin{align}
&\vartheta \left[
\begin{array}{c}
a\\ b
\end{array}
\right] (c(\nu +n),c\tau)
=e^{2\pi i acn}\vartheta \left[
\begin{array}{c}
a\\ b
\end{array}
\right] (c\nu,c\tau), \notag \\
&\vartheta \left[
\begin{array}{c}
a\\ b
\end{array}
\right] (c(\nu +n\tau),c\tau)
=e^{-i\pi cn^2\tau -2\pi i n(c\nu +b)}\vartheta \left[
\begin{array}{c}
a\\ b
\end{array}
\right] (c\nu,c\tau), \notag \\
&\vartheta \left[
\begin{array}{c}
a+m\\ b+n
\end{array}
\right] (c\nu,c\tau)
=e^{2\pi i an}\vartheta \left[
\begin{array}{c}
a\\ b
\end{array}
\right] (c\nu,c\tau),\notag \\
&\vartheta \left[
\begin{array}{c}
a\\ b
\end{array}
\right] (c\nu,c\tau)
	=
\vartheta \left[
\begin{array}{c}
a\\ 0
\end{array}
\right] (c\nu +b,c\tau),
	\label{thetafunction_properties}
\end{align}
where $a$ and $b$ are real numbers, $c$, $m$ and $n$ are integers, and $\nu$ and $\tau$ are complex numbers with $\mathrm{Im}\tau >0$.
The following orthonormality condition determines the normalization factor $\mathcal{N}_{|M_{ab}|}$,
\al{
\int_{T^2} d^2z
\left( \Theta^{(I + \alpha_{ab}, \beta_{ab})}_{M_{ab}} (z, \tau) \right)^\ast
\left( \Theta^{(J + \alpha_{ab}, \beta_{ab})}_{M_{ab}} (z, \tau) \right)
&= \delta_{I,J} \quad (M_{ab} > 0), \notag \\
\int_{T^2} d^2z
\left( \Theta^{(I + \alpha_{ba}, \beta_{ba})}_{M_{ba}} (\overline{z}, \overline{\tau}) \right)^\ast
\left( \Theta^{(J + \alpha_{ba}, \beta_{ba})}_{M_{ba}} (\overline{z}, \overline{\tau}) \right)
&= \delta_{I,J} \quad (M_{ba} < 0),
}
with $d^2z := dz d\overline{z}$.
An important relationship is easily derived (in the case of $M_{ab} > 0$),
\al{
\left( \Theta^{(I + \alpha_{ab}, \beta_{ab})}_{M_{ab}} (z, \tau) \right)^\ast
=
\Theta^{(-I + \alpha_{ba}, \beta_{ba})}_{M_{ba}} (\overline{z}, \overline{\tau}),
	\label{CC_theta}
}
where the index $I$ is identified under the condition, $\text{mod}\ |M_{ab}|$, and we can always redefine $-I$ as $I'(=0,\cdots, |M_{ab}|-1)$.

On {a} flux background, zero-mode profiles are not only split {but} also localized around {points different from} each other.
Then we can expect that hierarchical values in Yukawa couplings are created via overlap integrals in the Yukawa sector of this model.
The concrete form of the Yukawa couplings {is} as follows{,}
\al{
\lambda_{I, J, K} &= \int_{T^2} d^2z \,
	\Theta^{(I + \alpha_{I}, \beta_{I})}_{M_{I}} (z, \tau)
	\Theta^{(J + \alpha_{J}, \beta_{J})}_{M_{J}} (z, \tau)
	\left(
	\Theta^{(K + \alpha_{K}, \beta_{K})}_{M_{K}} (z, \tau)
	\right)^\ast,
	\label{i-th_torus_Yukawa}
}
where we drop a constant factor via gauge structure {and the indices $I,J,K$ discriminate degenerated states of three kinds of fields.
$M_{i}$, $\alpha_{i}$, $\beta_{i}$ ($i=I,J,K$) represent corresponding magnetic fluxes, two kinds of Scherk-Schwarz phases.}
In a suitable symmetry breaking like the above examples, we can find conditions on the parameters,
\al{
M_{I} + M_{J} &= M_{K}, \label{selectionrune_M} \\
\alpha_{I} + \alpha_{J} &= \alpha_{K}, \label{selectionrune_alpha} \\
\beta_{I} + \beta_{J} &= \beta_{K}, \label{selectionrune_beta}
}
where we implicitly use the rule in Eq.~(\ref{selectionrune_M}) and the relation in Eq.~(\ref{CC_theta}) when we write down the actual form in Eq.~(\ref{i-th_torus_Yukawa}).

After some mathematical calculations, we can derive the analytical result of the Yukawa coupling in Eq.~(\ref{i-th_torus_Yukawa}) as
\al{
\lambda_{I, J, K} &= \frac{\mathcal{N}_{M_I} \mathcal{N}_{M_J}}{\mathcal{N}_{M_K}}
\times \sum_{m \in Z_{M_K}}
\vartheta \left[
\begin{array}{c}
\frac{M_J \left(I + \alpha_I\right) - M_I \left(J + \alpha_J\right) + m M_I M_J}{M_I M_J M_K} \\ 0
\end{array}
\right] {(X, Y)} \notag \\
&\quad \times
	\delta_{I + \alpha_I + J + \alpha_J + m M_I, \, K + \alpha_K + \ell M_K},
	\label{general_Yukawa_formula}
}
with $X := M_I \beta_J - M_J \beta_I$, $Y := \tau M_I M_J M_K$ and possible choices of integers $\ell$~{\cite{Cremades:2004wa,Abe:2015yva}}.
In {other words}, we consider the Kronecker's delta with the condition {``$\text{mod}\ |M_{K}|$.''}

\subsection{Yukawa coupling on magnetized $T^{2}/Z_{N}$}

Next we consider the case on magnetized $T^{2}/Z_{N}$.
Here, the form of the fermion {wave function} with
the $Z_N$ parity $\eta$, the ({two-dimensional}) positive chirality and the state-discriminating index $I$ on $T^2$ is
\al{
\psi_{+,\eta}(z) = \sum_{I=0}^{|M| - 1} \psi^{I}_{+,\eta}(z),\quad
\psi^{I}_{+,\eta}(z) =
\begin{pmatrix}
\widetilde{\Theta}^{(I+\alpha, \beta)}_{M;\,\eta} (z, \tau) \\ 0
\end{pmatrix},
}
where we assume that $M$ is a positive integer.\footnote{
Note that the correspondence to the negative chirality case is basically straightforward by the replacements $z \to \overline{z}$, $\tau \to \overline{\tau}$.
}
Constructing the concrete form of $\widetilde{\Theta}^{(I+\alpha, \beta)}_{M;\, \eta} (z, \tau)$ itself can be done straightforwardly just following the general recipe as\footnote{Indeed, we have degrees of freedom of putting any terms on orbifold fixed points. However, for simplicity, we assume the absence of such terms.}
\al{
\widetilde{\Theta}^{(I+\alpha, \beta)}_{M;\, \eta} (z, \tau) =
\frac{1}{N} \sum_{x=0}^{N-1} \left(\overline{\eta}\right)^x
\Theta^{(I+\alpha, \beta)}_{M} (\omega^x z, \tau).
	\label{ZN_rotation}
}
Naively, Yukawa couplings on $T^{2}/Z_{N}$ seem to be formulated as 
\al{
\widetilde{\lambda}_{I, J, K} &= \int_{T^2} d^2z \,
	\widetilde{\Theta}^{(I + \alpha_{I}, \beta_{I})}_{M_{I};\, \eta_{I}} (z, \tau)
	\widetilde{\Theta}^{(J + \alpha_{J}, \beta_{J})}_{M_{J};\, \eta_{J}} (z, \tau)
	\left(
	\widetilde{\Theta}^{(K + \alpha_{K}, \beta_{K})}_{M_{K};\, \eta_{K}} (z, \tau)
	\right)^\ast,
	\label{T2ZN_Yukawa}
}
where we find the condition on the $Z_N$ parities (via the invariance of the system),\footnote{
Note that a complex-conjugated state {holds} the corresponding complex-conjugated $Z_N$ parity.
}
\al{
\eta_{I} \eta_{J} \overline{\eta_{K}} = 1.
}

However, this is not the end of the story.
In general, the kinetic terms on $T^2/Z_N$, which are described as
\al{
\mathcal{K}_{I J}^{(Z_N;\, \eta)} =
\int_{T^2} d^2z
\left( \widetilde{\Theta}^{(I+\alpha, \beta)}_{M;\, \eta} (z, \tau) \right)^\ast
\widetilde{\Theta}^{(J+\alpha, \beta)}_{M;\, \eta} (z, \tau),
}
are no longer diagonal, where the number of independent physical states should be reduced as ${\text{rank}\left[ \mathcal{K}_{I J}^{(Z_N;\, \eta)} \right]} < |M|$.
{Thereby,} in the physical eigenstates after considering the correct normalization in the kinetic terms by the unitary transformation with the corresponding diagonalizing matrix $U^{(Z_N;\, \eta)}$, where $\mathcal{K}_{I J}^{(Z_N;\, \eta)}$ is {transformed} as
\al{
\mathcal{K}^{(Z_N;\, \eta)} \to \left(U^{(Z_N;\, \eta)}\right)^\dagger \mathcal{K}^{(Z_N;\, \eta)} U^{(Z_N;\, \eta)}
=
\text{diag}(\underbrace{1,\cdots,1}_{{\text{Rank}\left[ \mathcal{K}^{(Z_N;\, \eta)} \right]}}, 0,\cdots,0),
	\label{diagonalizing_K}
}
{The mode} function on $T^2/Z_N$ should be
\al{
\widetilde{\Theta}^{(I+\alpha, \beta)}_{M;\, \eta} (z, \tau) \to
\sum_{I = 0}^{|M|-1}
\widetilde{\Theta}^{(I+\alpha, \beta)}_{M;\, \eta} (z, \tau)
\left(U^{(Z_N;\, \eta)}\right)_{I I'},
	\label{to_physicalbasis}
}
where $I'$ is the index of physical eigenstates from zero to ${\text{rank}\left[ \mathcal{K}_{I J}^{(Z_N;\, \eta)} \right]} -1$.
{The operator} formalism helps us to evaluate explicit forms of $\widetilde{\lambda}_{I, J, K}$ defined in Eq.~(\ref{T2ZN_Yukawa}) and the matrix $U^{(Z_N;\, \eta)}$ (see~\cite{Abe:2014noa,Abe:2015yva} for details.).

Taking into account {the} effects of the diagonalization, the final form of the Yukawa coupling is expressed as
\al{
\widetilde{\lambda}'_{I',J',K'} = \sum_{I=0}^{|M_I|-1} \sum_{J=0}^{|M_J|-1} \sum_{K=0}^{|M_K|-1}
\widetilde{\lambda}_{I, J, K} \left(U^{Z_N; \eta_I}\right)_{I,I'} \left(U^{Z_N; {\eta_J}}\right)_{J,J'}
\left({U^{Z_N; {\eta_K}}}\right)_{K,K'}^\ast
	\label{eq:final_formula_Yukawa}
}
where the indices for identifying kinetic eigenstates, $I'$, $J'$, $K'$, have {$\text{rank}\left[ \mathcal{K}^{(Z_N; \eta_I)} \right]$, $\text{rank}\left[ \mathcal{K}^{(Z_N; \eta_J)} \right]$, $\text{rank}\left[ \mathcal{K}^{(Z_N; \eta_K)} \right]$} numbers of nonzero configurations, respectively. 
In general, the mixing effect through $U^{Z_N; \eta_I}$ contributes to the physics.

\subsection{Possible configurations of three generations in $U(8)$ model}

We focus on the following pattern of the gauge symmetry breaking under the magnetic flux is $U(N) \to U(N_a) \times U(N_b) \times U(N_c)$ with $N = N_a + N_b + N_c$, where the corresponding 1-form potential is
\al{
&A^{(b)}(z, \overline{z}) = \frac{\pi}{q \text{Im} \tau} \times
\text{diag}\left(
M_a \text{Im}\left[ ( \overline{z} ) dz \right] \mathbf{1}_{N_a \times N_a},
M_b \text{Im}\left[ ( \overline{z} ) dz \right] \mathbf{1}_{N_b \times N_b},
M_c \text{Im}\left[ ( \overline{z} ) dz \right] \mathbf{1}_{N_c \times N_c}\right).
}
{We find six types of {bifundamental} matter fields under $U(N_a) \times U(N_b) \times U(N_c)$,
$\lambda^{ab}$,
$\lambda^{bc}$,
$\lambda^{ca}$,
$\lambda^{ba}$,
$\lambda^{cb}$,
$\lambda^{ac}$, whose gauge properties are
$(N_a, \overline{N_b}, 1)$,
$(1, N_b, \overline{N_c})$, 
$(\overline{N_a}, 1, N_c)$, 
$(\overline{N_a}, N_b, 1)$, 
$(1, \overline{N_b}, N_c)$, 
$(N_a, 1, \overline{N_c})$, respectively.}
When we adopt the choice $N_a = 4$, $N_b = 2$, $N_c = 2$, $U(4)_{PSC} \times U(2)_L \times U(2)_R$ gauge groups are realized from the $U(8)$ group up to $U(1)$ factors, where the subscripts $PSC$, $L$ and $R$ denote the Pati-Salam color, left- and right-electroweak gauge groups,\footnote{
{From} a phenomenological point of view, we can consider the following additional breakdowns originating from flux, $U(4)_{PSC} \to U(3)_C \times U(1)_1$ and $U(2)_R \to U(1)_2 \times U(1)_3$ (up to $U(1)$ factors), where $U(3)_C$ is the color gauge group (up to {a $U(1)$ factor}).
Under the latter breaking, the up-type and down-type Higgsino/Higgs sectors can feel different magnetic fluxes individually.
Consequently, the numbers of the two types of {fields diverge}.
Some of the combinations of the $U(1)$ part would be {anomalous. Then they could} be massive and decoupled via the Green-Schwarz mechanism.
}
respectively.
In such a situation, when the actual chirality of the gaugino is left (negative), {$\lambda^{ab}$ corresponds to the left-handed quarks and leptons, and $\lambda^{ca}$ accords with (charge-conjugated) right-handed quarks and leptons, respectively}.
When the magnetic fluxes are suitably assigned, the situation with {three generations} is materialized.
Besides, $\lambda^{bc}$ plays as up-type and down-type Higgsinos.
After we assume that ({four-dimensional} $\mathcal{N}=1$) supersymmetry is preserved at least locally at the $ab$, $bc$ and $ca$ sectors, the corresponding Higgses via extra-dimensional components of the {ten-dimensional} vector fields {are} still massless under the fluxes and the number of the fields are the same with Higgsino fields.
Also, no tachyonic mode is expected at the tree level.
Here, in general, multiple Higgs fields appear from the $bc$ sector.
Interestingly, when $\lambda^{ab}$, $\lambda^{bc}$ and $\lambda^{ca}$ have zero modes, $\lambda^{ba}$, $\lambda^{cb}$ and $\lambda^{ac}$ cannot contain any zero mode and thus no exotic particle arises from these fermionic sectors.
In the case of the actual chirality being right (positive), we should flip the roles of the two categories.
The $U(8)$ gauge group is the minimal group for matter unification within the $U(N)$ gauge theories.\footnote{In the previous models \cite{Abe:2012fj, Abe:2014vza}, flavor structures among quarks and leptons are characterized only on two dimensions of the {six-dimensional} compact space. In this paper, we focus on a two-dimensional toroidal orbifold{,} and our setup {is} expected to be embedded in $SO(32)$ SYM theory and {six-dimensional}/{ten-dimensional} other theories with $U(N)$ groups.}
The following properties are observed,
\al{
M_{ab} + M_{bc} + M_{ca} &= 0, \notag \\
\alpha_{ab} + \alpha_{bc} + \alpha_{ca} &= 0, \notag \\
\beta_{ab} + \beta_{bc} + \beta_{ca} &= 0, \notag \\
\eta_{ab} \eta_{bc} \eta_{ca} &= 1
	\label{constraints_on_parameters}
}
where the above parameters are defined by the fundamental ones like $M_{ab} = M_{a} - M_{b}$ except the $Z_N$ parities.
The $Z_N$ parities are described as
$
\eta_{ab} = \eta_{a} \overline{\eta_{b}},\
\eta_{bc} = \eta_{b} \overline{\eta_{c}},\
\eta_{ca} = \eta_{c} \overline{\eta_{a}}.
$

We assume both {nonvanishing} magnetic fluxes and orbifold twists. Indeed, thanks to magnetic fluxes, there is the possibility that ${\cal N}=4$ SUSY in {four-dimensional} spacetime is broken into ${\cal N}=1, 2,$ or $0$ (non-SUSY case). Similarly, orbifold twists can break ${\cal N}=4$ SUSY into $N=1$ or $2$. In this paper, we do not try to construct concrete full setups. However, we can construct $N = 1$ SUSY models by selecting magnetic fluxes and orbifold twists suitably.\footnote{In order to realize $N=1$ SUSY models, we need to appropriately assign magnetic fluxes and/or boundary conditions of orbifolding. See, e.g., Ref.~\cite{Abe:2012fj}.}

All the possible configurations {with three generations} which fulfill the conditions in Eq.~(\ref{constraints_on_parameters}) were derived in Ref.~\cite{Abe:2015yva}.\footnote{In our setup, {R-parity-violating} terms are prohibited by Pati-Salam and/or U(1) gauge symmetries. {For a review, see} Ref.~\cite{Abe:2012fj}}
{Note that the constraints $\alpha_{i} = \beta_{i}\,(i = ab,\, bc,\, ca)$ are requested by symmetry in the cases of $T^{2}/Z_{3,4,6}$.}
We note that the first line of Eq.~(\ref{constraints_on_parameters}) tells us that at least one of the signs of the three fluxes should be different from the others
and it is enough that we focus on the two possibilities in the signs of the fluxes of the matter sectors {$M_{ab}$ and {$M_{ca}$} as}
\al{
M_{ab} < 0,\ {M_{ca}} < 0; \quad
M_{ab} < 0,\ {M_{ca}} > 0.
	\label{scanningcondition_M_1}
}
There is another possibility of $M_{ab} > 0,\ {M_{ca}} < 0$, but this case is physically the same as $M_{ab} < 0,\ {M_{ca}} > 0$.
Besides, after ignoring the difference coming from the combinatorics, we can introduce the additional condition,
\al{
{|M_{ab}| \leq |{M_{ca}}|}.
	\label{scanningcondition_M_2}
}
Results of classification are shown in Tables~\ref{tbl:Z2_classification_result} ($Z_{2}$), \ref{tbl:Z3_classification_result} ($Z_{3}$), \ref{tbl:Z4_classification_result} ($Z_{4}$), \ref{tbl:Z6_classification_result} ($Z_{6}$), respectively.

\begin{table}[H]
\begin{center}
\begin{tabular}{|l|l||l|l|} \hline
\multicolumn{2}{|c||}{General cases} & \multicolumn{2}{|c|}{Trivial {BCs} only} \\ \hline
$M_{ab},M_{ca} < 0$ & $M_{ab} < 0,\, M_{ca} > 0$ & $M_{ab},M_{ca} < 0$ & $M_{ab} < 0,\, M_{ca} > 0$ \\ \hline \hline
$41\,(N_{H} = 5)$ & $16\,(N_{H} = 1_{\text{trivial}})$ & $5\,(N_{H} = 5)$ & $4\,(N_{H} = 1_{\text{trivial}})$ \\
$56\,(N_{H} = 6)$ & $65\,(N_{H} = 1)$ & $2\,(N_{H} = 6)$ & $5\,(N_{H} = 1)$ \\
$30\,(N_{H} = 7)$ &    &     &    \\
$8\,(N_{H} = 8)$ &     & $2\,(N_{H} = 8)$ &    \\
$1\,(N_{H} = 9)$ &     & $1\,(N_{H} = 9)$ &    \\ \hline
\multicolumn{2}{|c||}{$136+81=217$ in total} & \multicolumn{2}{|c|}{$10+9=19$ in total} \\ \hline
\end{tabular}
\caption{Numbers of possible configurations with three generations in {the} $Z_2$ case.
``General cases'' and ``Trivial {BCs} only'' means the cases with and without nontrivial Scherk-Schwarz phases, respectively.
Corresponding numbers of the Higgs pairs ($N_H$) are also shown.
The case indicated by $1_{\text{trivial}}$ means the one Higgs pair appears under {the nonmagnetized} background in {the} $bc$ sector.
}
\label{tbl:Z2_classification_result}
\end{center}
\end{table}

\begin{table}[H]
\begin{center}
\begin{tabular}{|l|l||l|l|} \hline
\multicolumn{2}{|c||}{General cases} & \multicolumn{2}{|c|}{Trivial {BCs} only} \\ \hline
$M_{ab},M_{ca} < 0$ & $M_{ab} < 0,\, M_{ca} > 0$ & $M_{ab},M_{ca} < 0$ & $M_{ab} < 0,\, M_{ca} > 0$ \\ \hline \hline
$11\,(N_{H} = 4)$ & $17\,(N_{H} = 1_{\text{trivial}})$ & $1\,(N_{H} = 4)$ & $9\,(N_{H} = 1_{\text{trivial}})$ \\
$83\,(N_{H} = 5)$ & $142\,(N_{H} = 1)$ & $6\,(N_{H} = 5)$ & $27\,(N_{H} = 1)$ \\
$190\,(N_{H} = 6)$ & $21\,(N_{H} = 2)$ & $7\,(N_{H} = 6)$ &    \\
$83\,(N_{H} = 7)$ &     & $6\,(N_{H} = 7)$ &    \\
$11\,(N_{H} = 8)$ &     & $1\,(N_{H} = 8)$ &    \\ \hline
\multicolumn{2}{|c||}{$378+180=558$ in total} & \multicolumn{2}{|c|}{$21+36=57$ in total} \\ \hline
\end{tabular}
\caption{{Numbers of possible configurations with three generations} in $Z_3$ case. The convention is the same {in} Table~\ref{tbl:Z2_classification_result}.}
\label{tbl:Z3_classification_result}
\end{center}
\end{table}

\begin{table}[H]
\begin{center}
\begin{tabular}{|l|l||l|l|} \hline
\multicolumn{2}{|c||}{General cases} & \multicolumn{2}{|c|}{Trivial {BCs} only} \\ \hline
$M_{ab},M_{ca} < 0$ & $M_{ab} < 0,\, M_{ca} > 0$ & $M_{ab},M_{ca} < 0$ & $M_{ab} < 0,\, M_{ca} > 0$ \\ \hline \hline
$9\,(N_{H} = 4)$ & $24\,(N_{H} = 1_{\text{trivial}})$ & $3\,(N_{H} = 4)$ & $12\,(N_{H} = 1_{\text{trivial}})$ \\
$128\,(N_{H} = 5)$ & $228\,(N_{H} = 1)$ & $37\,(N_{H} = 5)$ & $60\,(N_{H} = 1)$ \\
$254\,(N_{H} = 6)$ & $18\,(N_{H} = 2)$ & $59\,(N_{H} = 6)$ & $6\,(N_{H} = 2)$ \\
$120\,(N_{H} = 7)$ &     & $27\,(N_{H} = 7)$ &    \\
$17\,(N_{H} = 8)$ &     & $10\,(N_{H} = 8)$ &    \\ \hline
\multicolumn{2}{|c||}{$528+270=798$ in total} & \multicolumn{2}{|c|}{$136+78=214$ in total} \\ \hline
\end{tabular}
\caption{{Numbers of possible configurations with three generations} in $Z_4$ case. The convention is the same {in} Table~\ref{tbl:Z2_classification_result}.}
\label{tbl:Z4_classification_result}
\end{center}
\end{table}

\begin{table}[H]
\begin{center}
\begin{tabular}{|l|l||l|l|} \hline
\multicolumn{2}{|c||}{General cases} & \multicolumn{2}{|c|}{Trivial {BCs} only} \\ \hline
$M_{ab},M_{ca} < 0$ & $M_{ab} < 0,\, M_{ca} > 0$ & $M_{ab},M_{ca} < 0$ & $M_{ab} < 0,\, M_{ca} > 0$ \\ \hline \hline
$14\,(N_{H} = 4)$ & $24\,(N_{H} = 1_{\text{trivial}})$ & $4\,(N_{H} = 4)$ & $12\,(N_{H} = 1_{\text{trivial}})$ \\
$156\,(N_{H} = 5)$ & $282\,(N_{H} = 1)$ & $45\,(N_{H} = 5)$ & $73\,(N_{H} = 1)$ \\
$326\,(N_{H} = 6)$ & $27\,(N_{H} = 2)$ & $76\,(N_{H} = 6)$ & $8\,(N_{H} = 2)$ \\
$150\,(N_{H} = 7)$ &     & $36\,(N_{H} = 7)$ &    \\
$20\,(N_{H} = 8)$ &     & $10\,(N_{H} = 8)$ &    \\ \hline
\multicolumn{2}{|c||}{$666+333=999$ in total} & \multicolumn{2}{|c|}{$171+93=264$ in total} \\ \hline
\end{tabular}
\caption{{Numbers of possible configurations with three generations} in $Z_6$ case. The convention is the same {in} Table~\ref{tbl:Z2_classification_result}.}
\label{tbl:Z6_classification_result}
\end{center}
\end{table}

\section{Results
\label{sec:result}}

Now we know that possible numbers of $SU(2)_{L}$ doublet Higgs boson pairs for up-type and down-type fields under the presence of magnetic fluxes are one or two (when $M_{ab} < 0,\, M_{ca} > 0$) or from four to eight (when $M_{ab} < 0,\, M_{ca} < 0$), respectively.
In the latter case, it had been investigated that when a suitable relation was fulfilled among the VEVs of the Higgs bosons, {the} so-called Gaussian Frogatt-Nielsen mechanism works and observed quark mass hierarchies and mixing angles are 
realized~\cite{Abe:2014vza}.
On the other hand, such cases are less predictive in the sense that various additional parameters with respect to scalar VEVs are required.

The situation in the former case is just opposite.
Here, only {lower} numbers of VEVs contribute to Yukawa hierarchies and {are} then more predictive, whereas less degrees of freedom can be used for realizations of the quark and lepton configurations in the SM.
In this work, we only focus on the magnitude of realized fermion mass hierarchies in the former case to declare prospects in such simple possibilities exhaustively.
{We comment on the total numbers of configurations found in panels in Figs.~\ref{fig:result_oneHiggs_1}, \ref{fig:result_oneHiggs_2}, \ref{fig:result_twoHiggs_Z3}, \ref{fig:result_twoHiggs_Z4} and \ref{fig:result_twoHiggs_Z6} are in general less than the numbers of corresponding allowed configurations shown in Tables~\ref{tbl:Z2_classification_result}, \ref{tbl:Z3_classification_result}, \ref{tbl:Z4_classification_result} and \ref{tbl:Z6_classification_result}.}
This is because part of configurations in Tables~\ref{tbl:Z2_classification_result}, \ref{tbl:Z3_classification_result}, \ref{tbl:Z4_classification_result} and \ref{tbl:Z6_classification_result} {results} in mass matrices with rank reduction (less than three), where such cases are apparently not suitable and skipped to be shown.
{The} cancellations after summing up all the indices in Eq.~(\ref{eq:final_formula_Yukawa}) {are a possible origin of this reduction}.

\subsection{One Higgs (pair) case}

At first, we consider the case with one Higgs pair.
We note that except for $T^{2}/Z_{2}$, the modulus parameter is inevitably fixed by requirement in $Z_{N}\,(N=3,4,6)$ orbifolds as $\tau = e^{2\pi i/N}$.
In other words, one additional parameter exists only in $T^{2}/Z_{2}$, where the magnitude of $\tau$ determines {the degrees of quasilocalization of the} mode functions.

\begin{figure}[H]
\centering
\includegraphics[width=0.32\columnwidth]{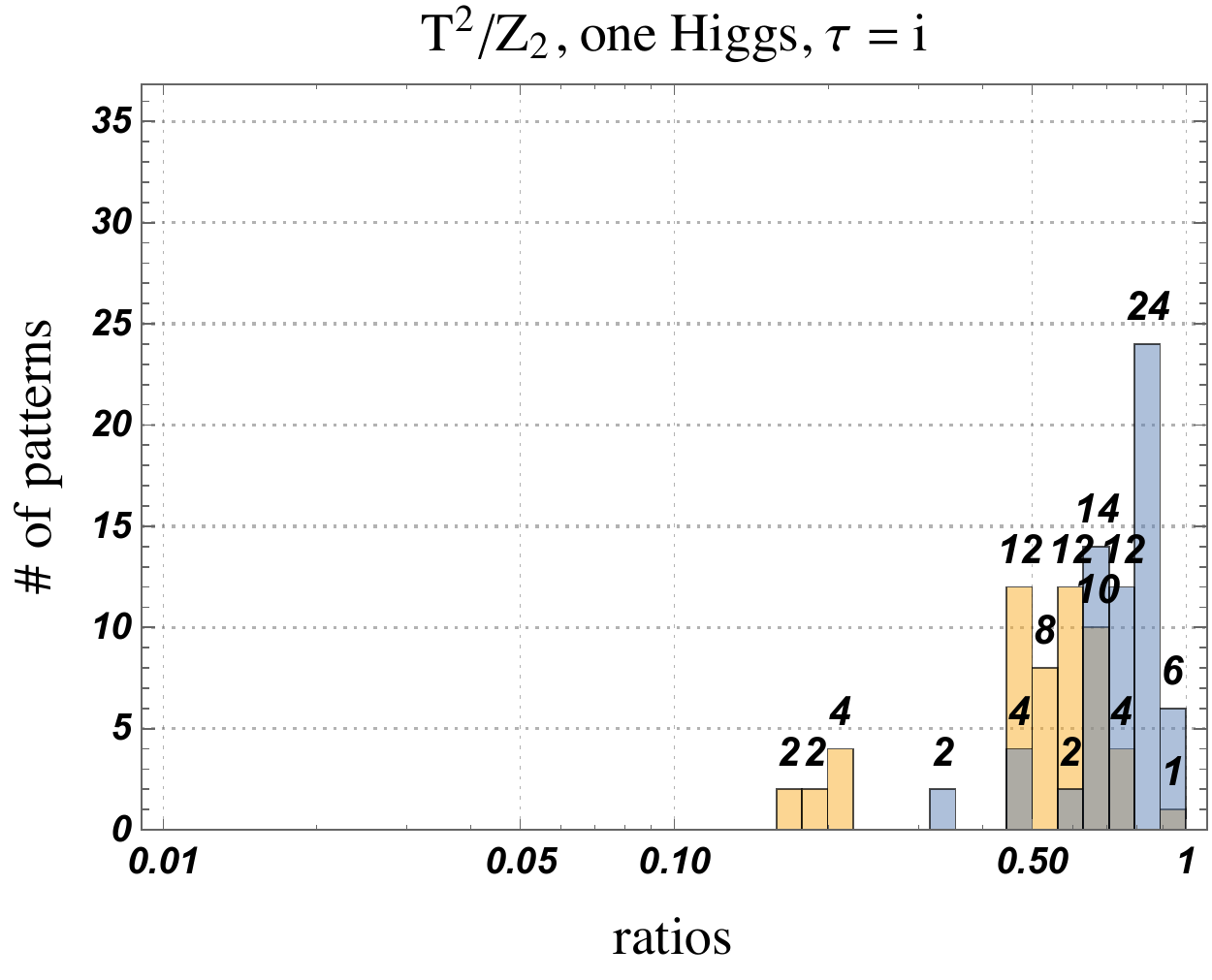}\ 
\includegraphics[width=0.32\columnwidth]{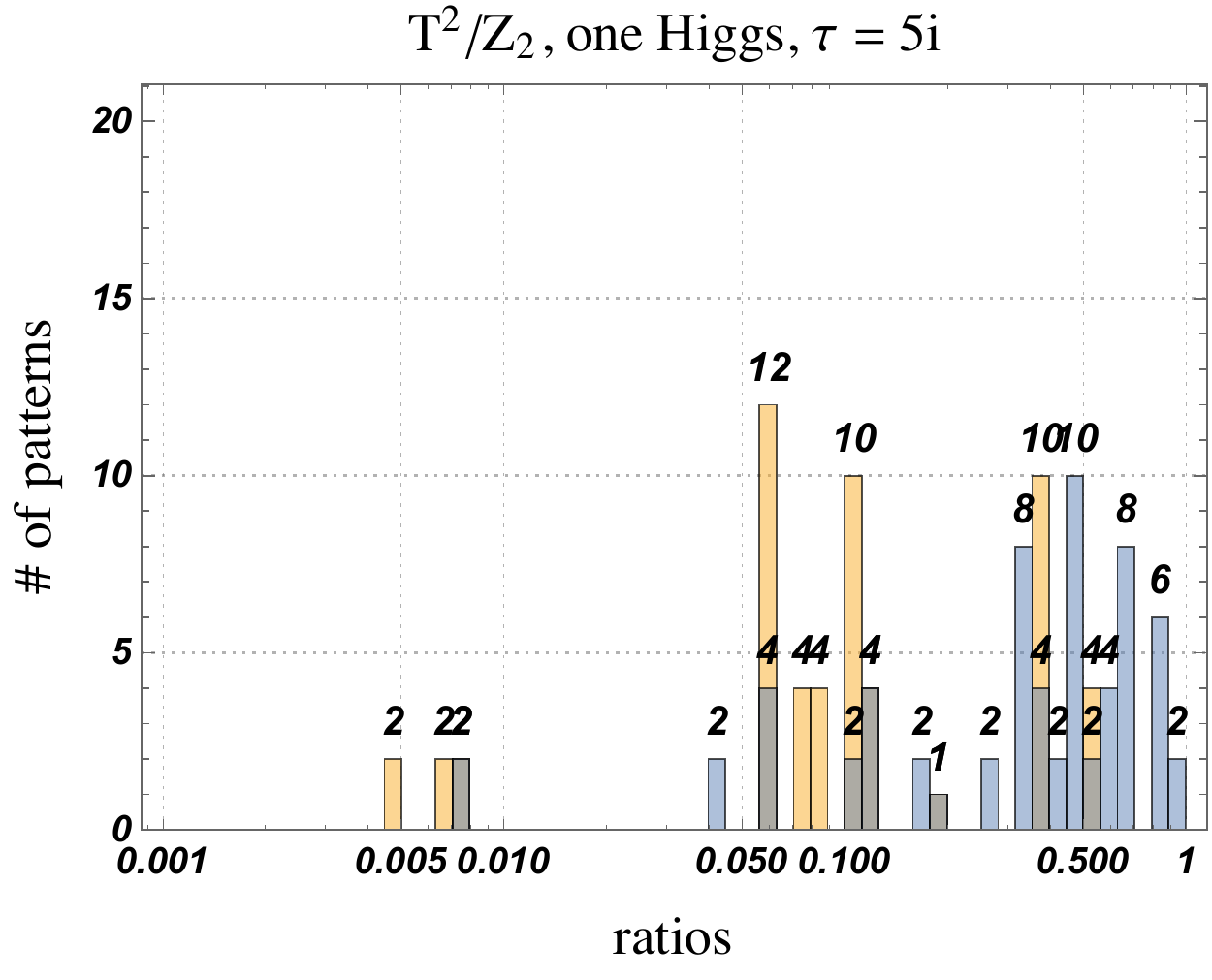}\
\includegraphics[width=0.32\columnwidth]{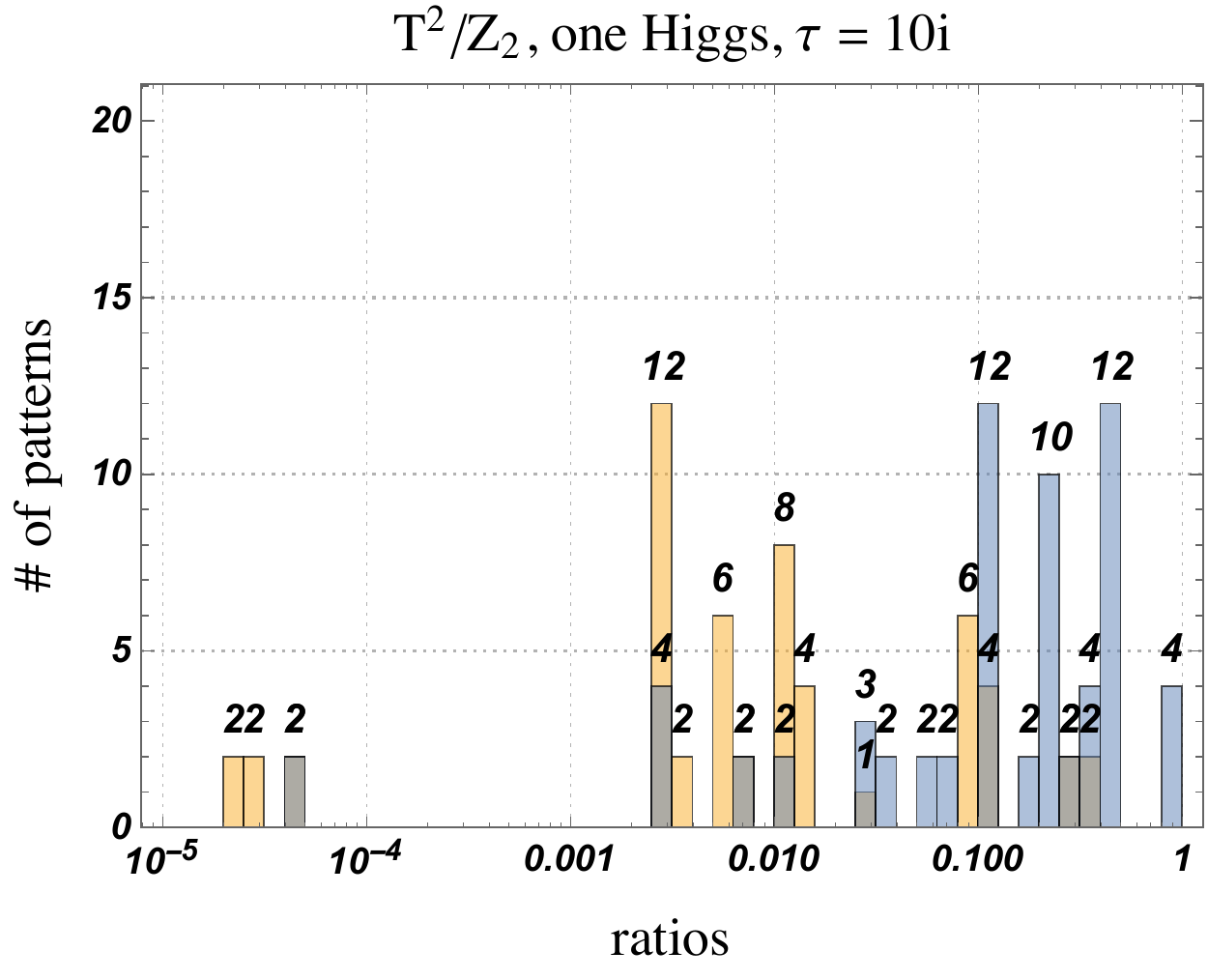}
\caption{
Distributions of realized mass eigenvalues are shown in $T^{2}/Z_{2}$ when one Higgs boson appears in the three choices of the modulus parameter $\tau = {i}$ ({left panel}), ${5\,i}$ ({center panel}), ${10\,i}$ ({right panel}).
The orange (blue) bars correspond to the mass ratio $m_{1}/m_{3}$ ($m_{2}/m_{3}$) under the ordering $m_{1} \leq m_{2} \leq m_{3}$.
{The total number of the possibilities with three generations is $65$ (in each panel).}
}
\label{fig:result_oneHiggs_1}
\end{figure}

\begin{figure}[H]
\centering
\includegraphics[width=0.32\columnwidth]{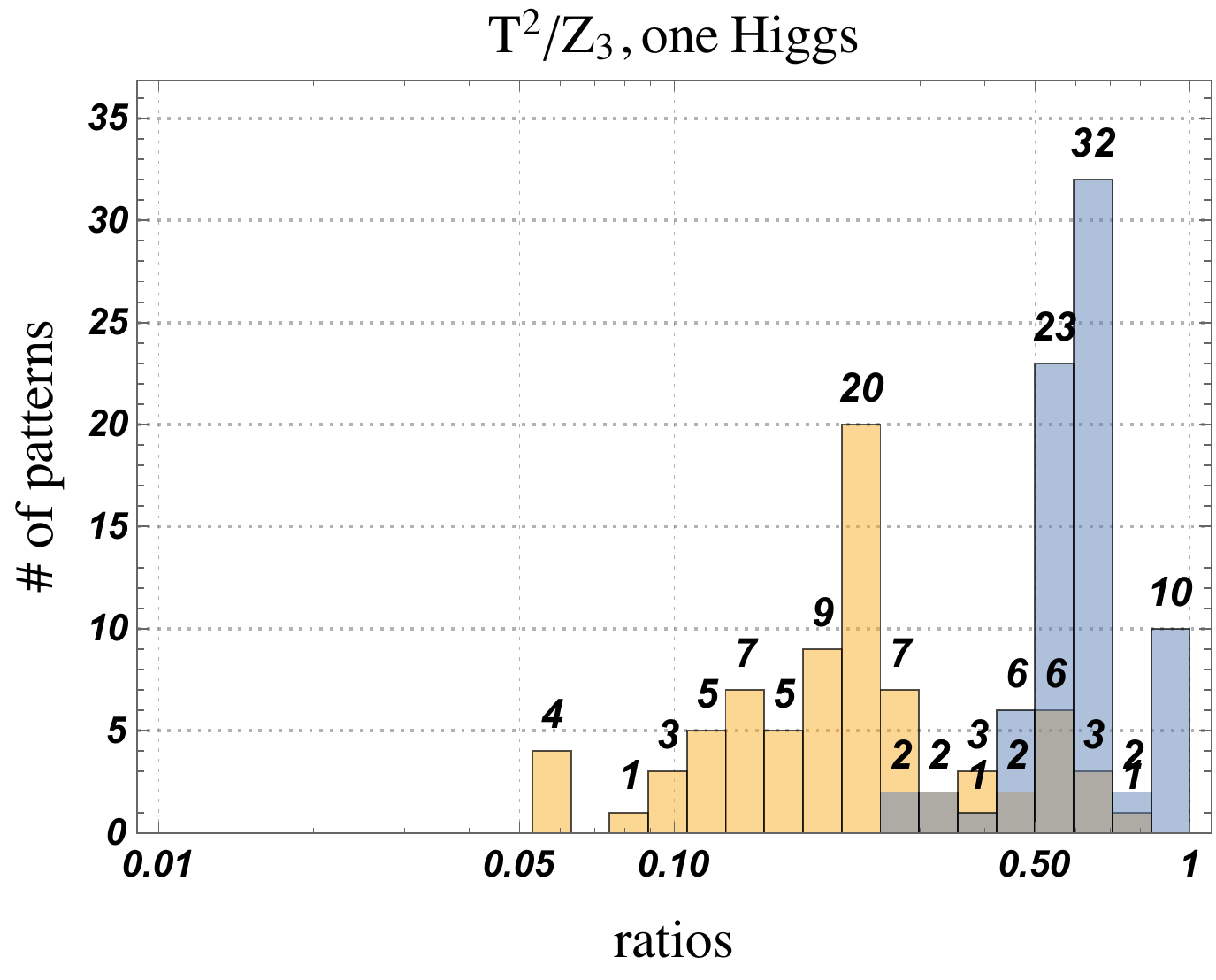}\ 
\includegraphics[width=0.32\columnwidth]{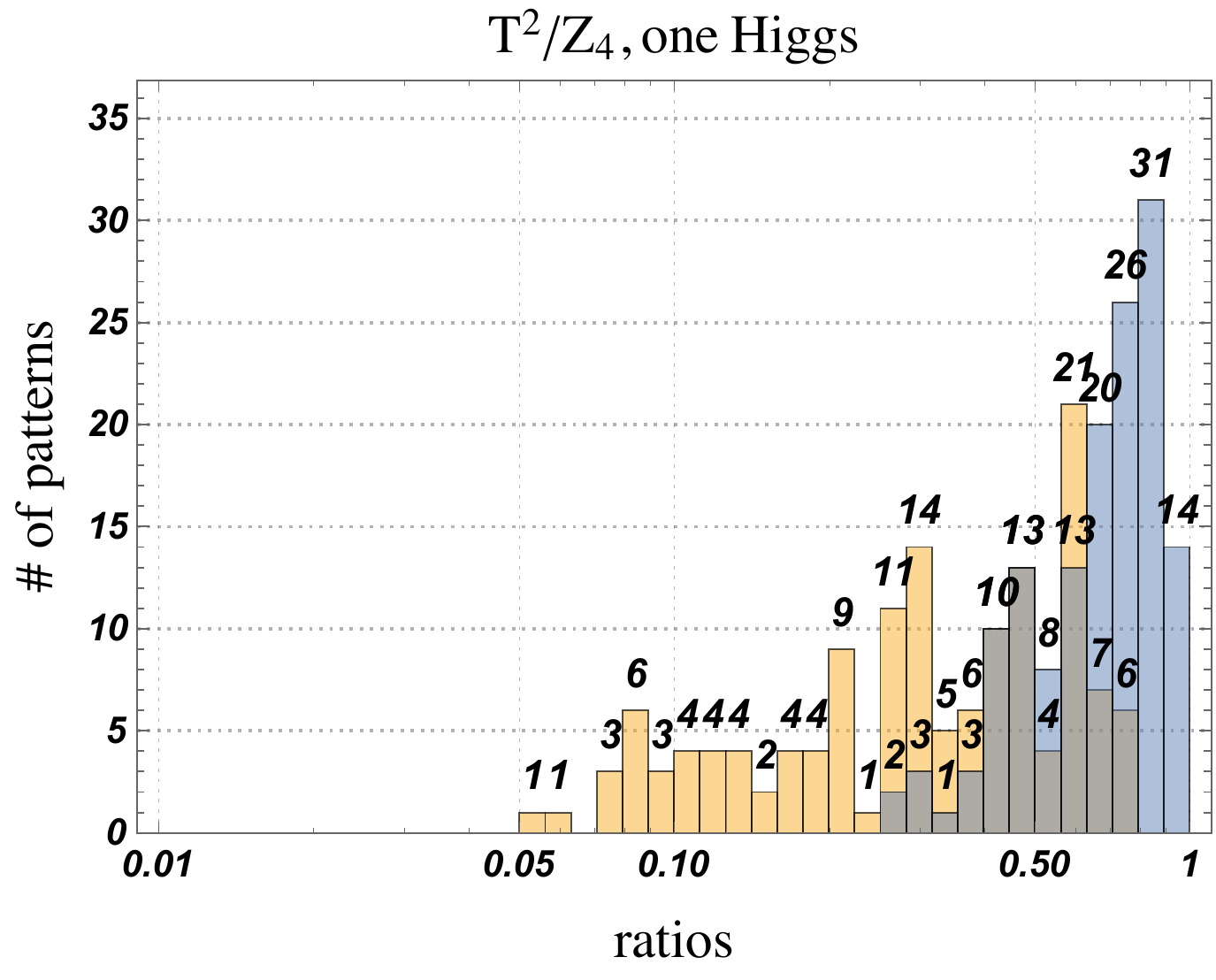}\
\includegraphics[width=0.32\columnwidth]{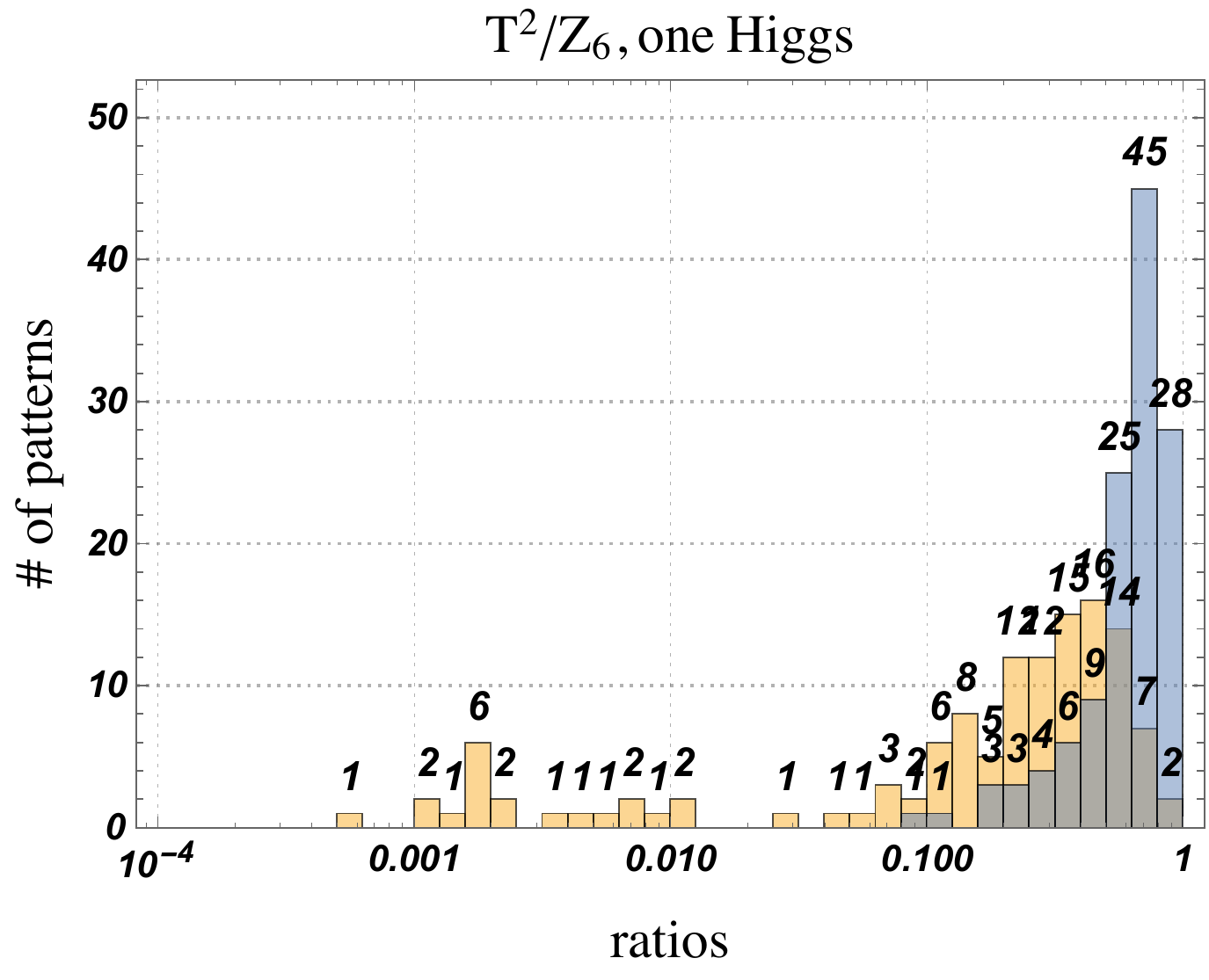}
\caption{
Distributions of realized mass eigenvalues are shown when one Higgs boson appears in the cases of $T^{2}/Z_{3}\,(\text{left panel}), T^{2}/Z_{4}\,(\text{center panel}), T^{2}/Z_{6}\,(\text{right panel})$.
The orange (blue) bars correspond to the mass ratio $m_{1}/m_{3}$ ($m_{2}/m_{3}$) under the ordering $m_{1} \leq m_{2} \leq m_{3}$.
{The total numbers of the possibilities with three generations are $78\,(T^{2}/Z_{3})$, $144\,(T^{2}/Z_{4})$, $135\,(T^{2}/Z_{6})$, respectively.}
}
\label{fig:result_oneHiggs_2}
\end{figure}

In Fig.~\ref{fig:result_oneHiggs_1}, distributions of realized mass eigenvalues are shown in $T^{2}/Z_{2}$ when one Higgs boson appears in the three choices of the modulus parameter $\tau = {i}$ ({left panel}), ${5\,i}$ ({center panel}), ${10\,i}$ ({right panel}).
Here, the orange (blue) bars correspond to the mass ratio $m_{1}/m_{3}$ ($m_{2}/m_{3}$) under the ordering $m_{1} \leq m_{2} \leq m_{3}$.
Digits on top of bars indicate {how many} configurations are stored in corresponding regions of $m_{1}/m_{3}$ or $m_{2}/m_{3}$.
Our result is consistent with those in the previous studies, e.g., in~\cite{Abe:2008fi,Abe:2008sx,Abe:2015yva}.
When we take {$\text{Im}[\tau]$} as greater than around ten, the hierarchy of $10^{-5}$ is realizable in $m_{1}/m_{3}$.
Then, we can realize the mass difference between the up quark and the top quark in this class by choosing the value of {$\text{Im}[\tau]$} as ten or a bit more.

In Fig.~\ref{fig:result_oneHiggs_2}, counterparts are depicted when one Higgs boson appears in the cases of $T^{2}/Z_{3}$ ({left panel}), $T^{2}/Z_{4}$ ({center panel}), $T^{2}/Z_{6}$ ({right panel}).
Conventions are the same as those in Fig.~\ref{fig:result_oneHiggs_1}.
Note that no additional parameters are there in these cases and then possibilities with {one Higgs} pair on $T^{2}/Z_{3,4,6}$ are completed in the figures.
Unfortunately, the minima of $m_{1}/m_{3}$ are around $5 \times 10^{-2}$ (for $T^{2}/Z_{3,4}$) and $1 \times10^{-3}$ (for $T^{2}/Z_{6}$), where {they} are far from the target $10^{-5}$.
Thereby, the simplest case in the number of realized Higgs pairs ($N_{H} = 1$) is discarded on $T^{2}/Z_{3,4,6}$ completely as candidates for describing the SM fermion sector.
Such {a} tendency is understandable via the formula in Eq.~(\ref{eq:final_formula_Yukawa}).
Different from the $Z_{2}$ case, in more than {the} $Z_{3}$ cases, {the} structure of the kinetic mixing described {by} the unitary matrix $U^{Z_{N};\,\eta}$ becomes nontrivial and such mixings tend to smear the difference in the original $T^{2}$ basis, where at this stage, large hierarchies are expected since the magnitude of magnetic fluxes {is} larger compared with when the geometry is $T^{2}/Z_{2}$ or the simple $T^{2}$.
In the case of $Z_{6}$, the highest magnitudes in magnetic fluxes are realized, where a bit more hierarchy is expected in spite of the smearing through the kinetic mixing.
This would be the origin {of} why the minimum of $m_{1}/m_{3}$ is a bit smaller on $T^{2}/Z_{6}$ than on $T^{2}/Z_{3,4}$.
Note that a similar conclusion was made in Ref.~\cite{Matsumoto:2016okl}.

\subsection{Two Higgs (pair) case}

Next, we go for the case with two Higgs pairs ($N_{H} = 2$), where two Higgs VEVs {contribute} to ratios of mass eigenvalues.
As far as we focus on the ratios of mass eigenvalues, only the ratio of the two VEVs {$v_{1}/v_{2}$} [$v_{1,2}$ corresponding to $J'=0,1$ in Eq.~(\ref{eq:final_formula_Yukawa})] is relevant.\footnote{We may suffer from large {flavor-changing neutral currents} (FCNCs) because two or more than two Higgs doublets emerge unless only one pair of Higgs doublets {stays} in the electroweak scale, while the others are decoupled. Even if sufficient Yukawa hierarchies are realized in our setup, we have to assume the Higgs sector including a specific Higgs mass matrix that can provide a light mass eigenstate and sufficiently heavy mass eigenstates of Higgs doublets {for} evading the FCNC effects.
Also, it is still a challenging issue to obtain such a Higgs sector.}

We find that nine, six and seven configurations generate rank-three mass matrices on $T^{2}/Z_{3}$, $T^{2}/Z_{4}$ and $T^{2}/Z_{6}$, respectively.
In each case, we examine patterns of $m_{1}/m_{3}$ and $m_{2}/m_{3}$ by plotting the ratios in various choices of $v_{1}/v_{2}$.
Note that no $N_{H} = 2$ example exists when the background is $T^{2}/Z_{2}$ as explicitly shown in Table~\ref{tbl:Z2_classification_result}.

\begin{figure}[H]
\centering
\includegraphics[width=0.32\columnwidth]{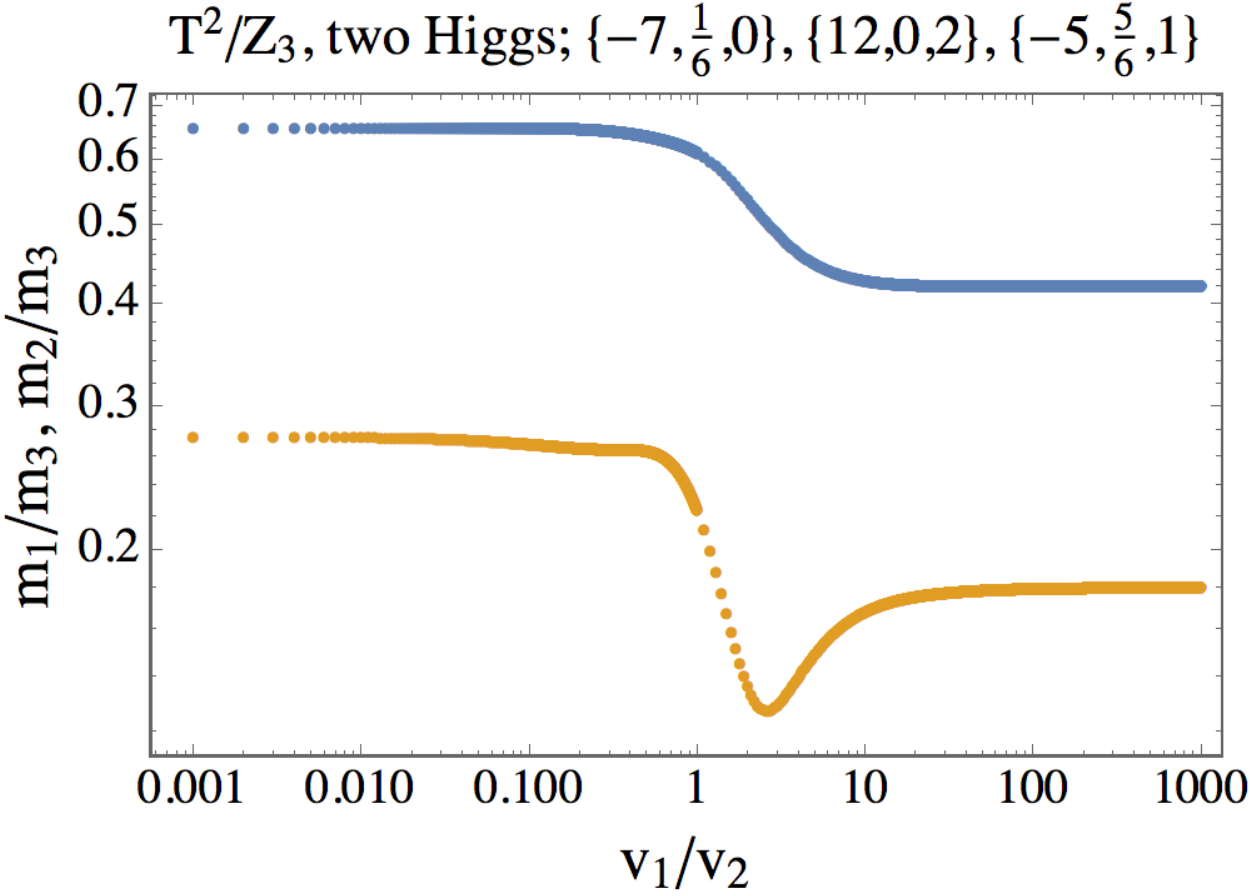}\ 
\includegraphics[width=0.32\columnwidth]{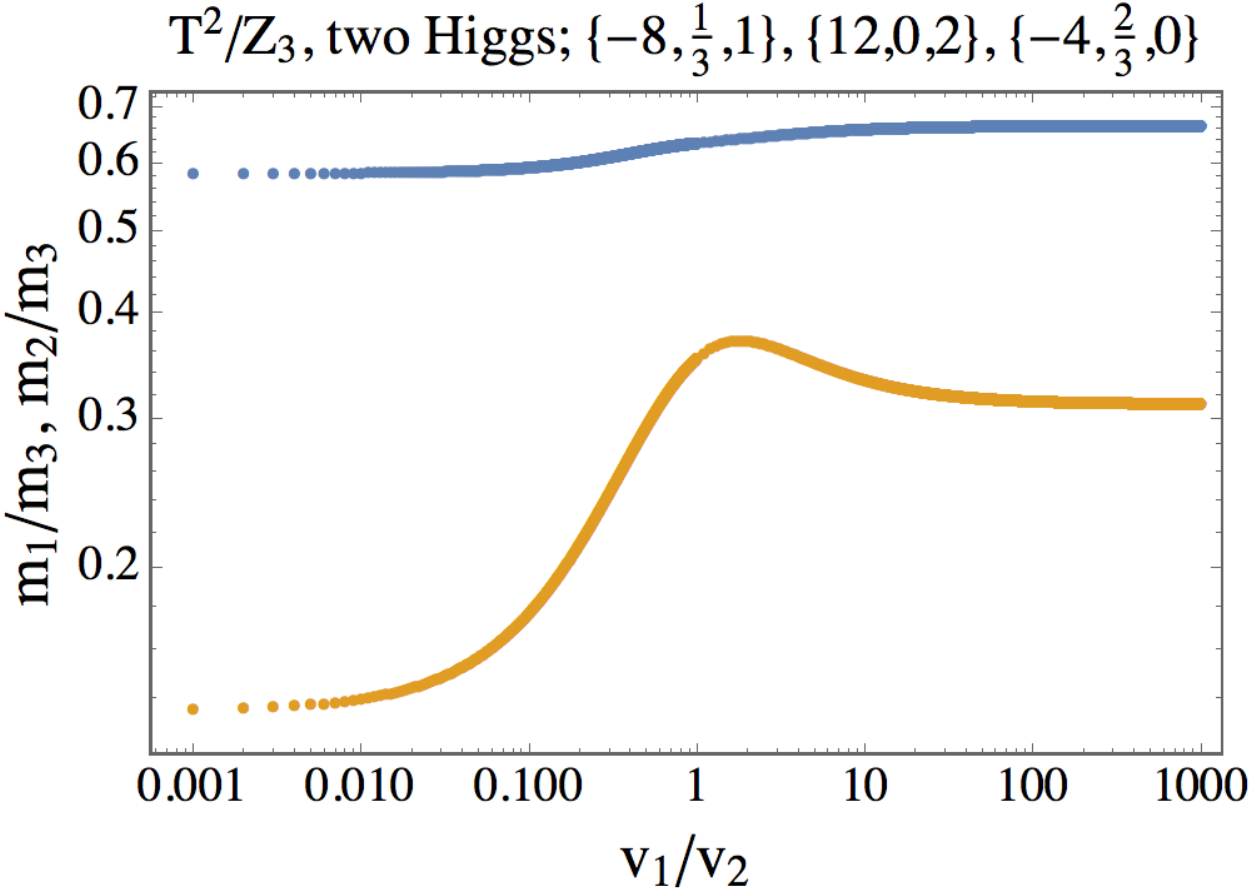}\
\includegraphics[width=0.32\columnwidth]{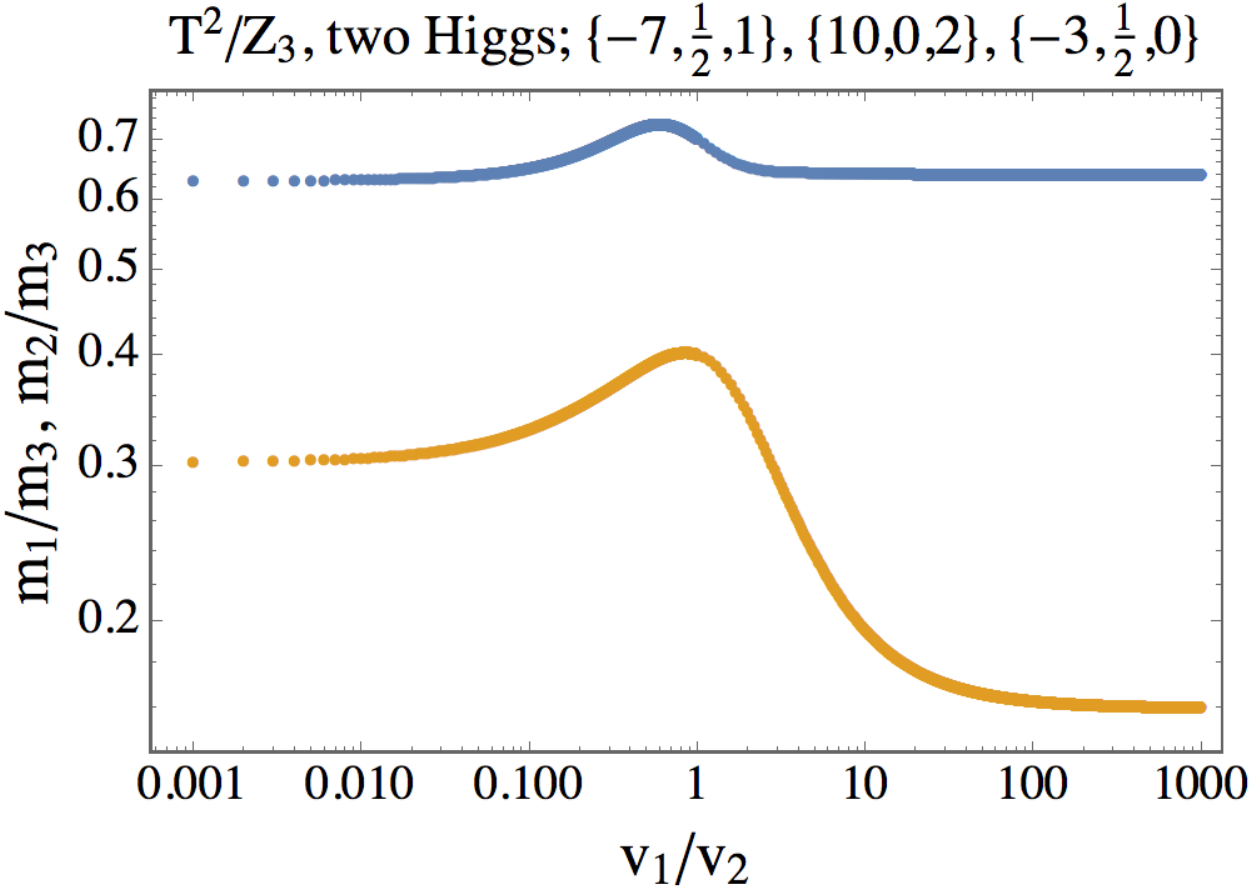} \\[6pt]
\includegraphics[width=0.32\columnwidth]{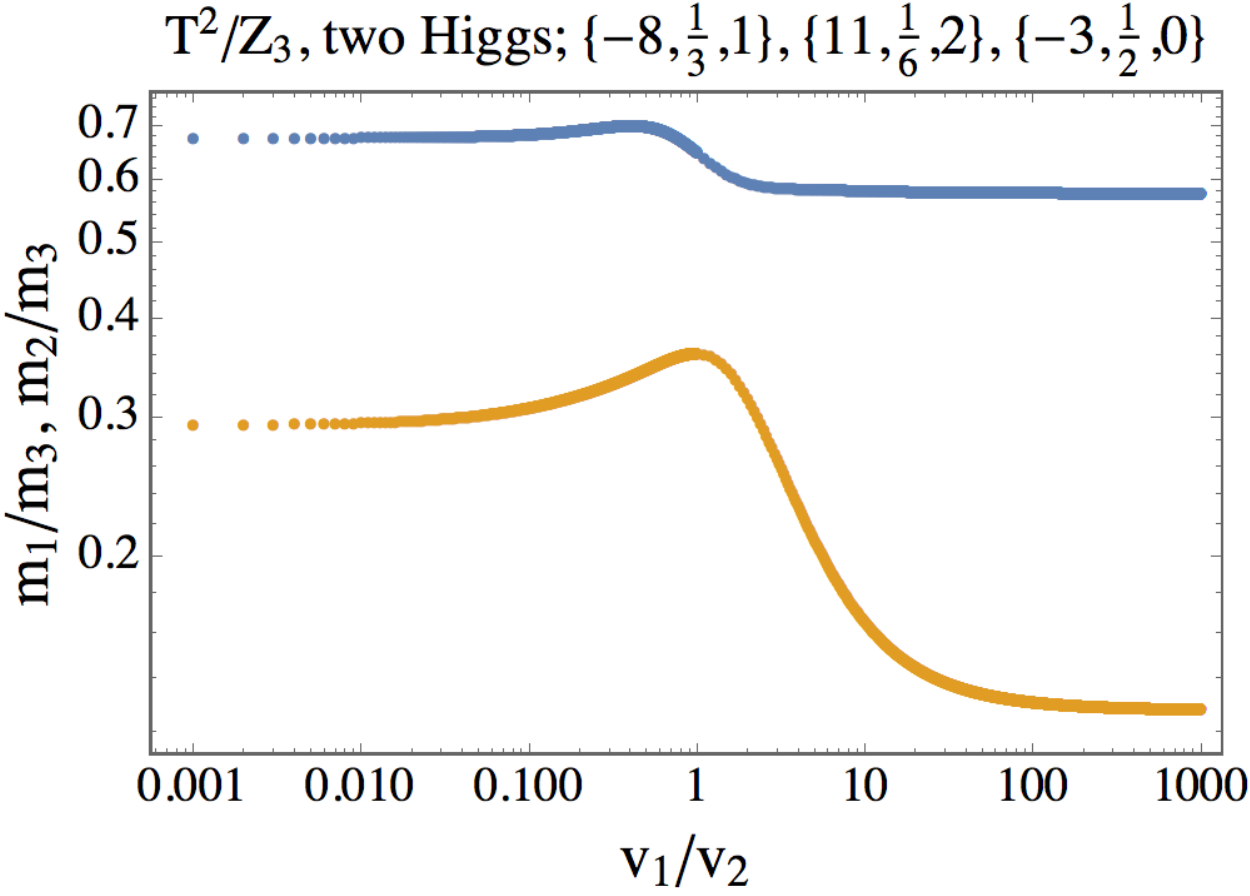}\ 
\includegraphics[width=0.32\columnwidth]{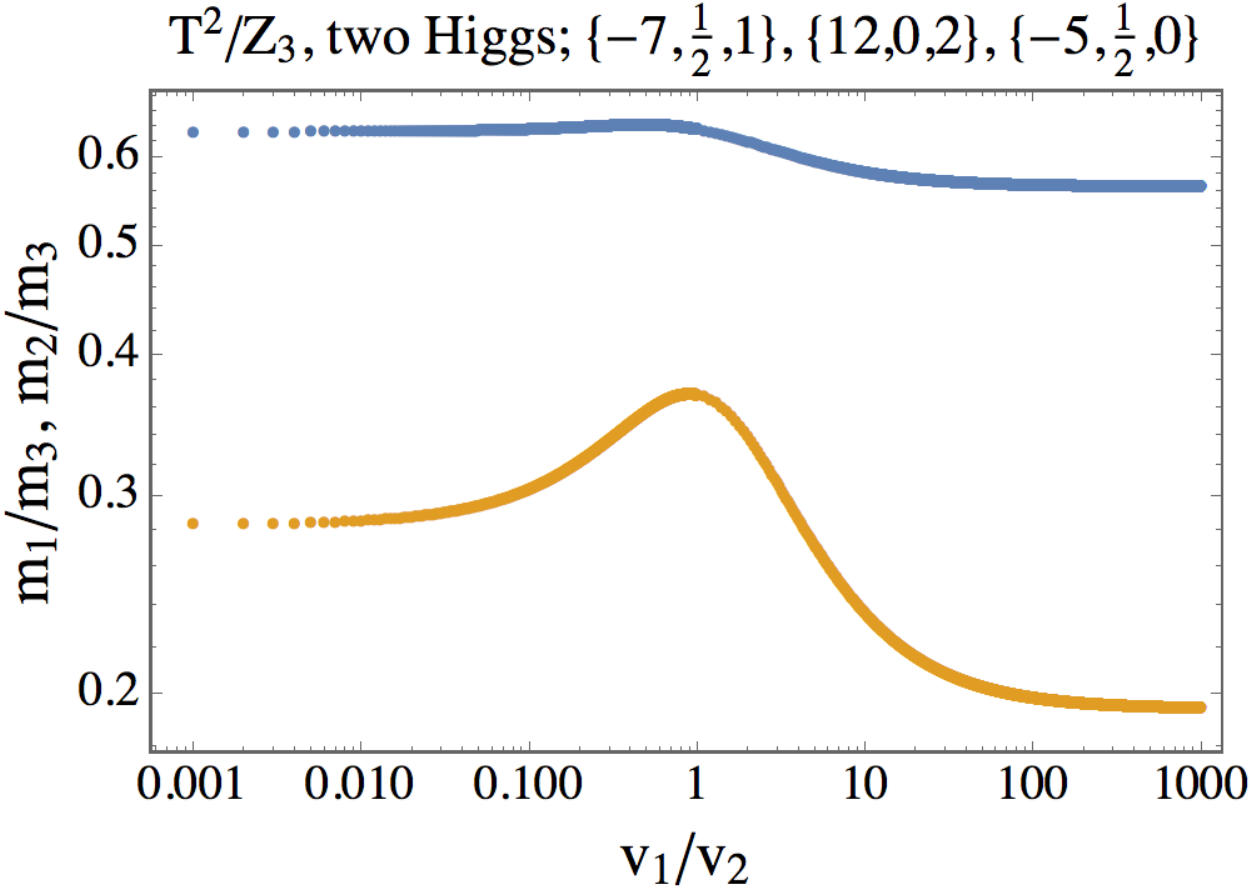}\
\includegraphics[width=0.32\columnwidth]{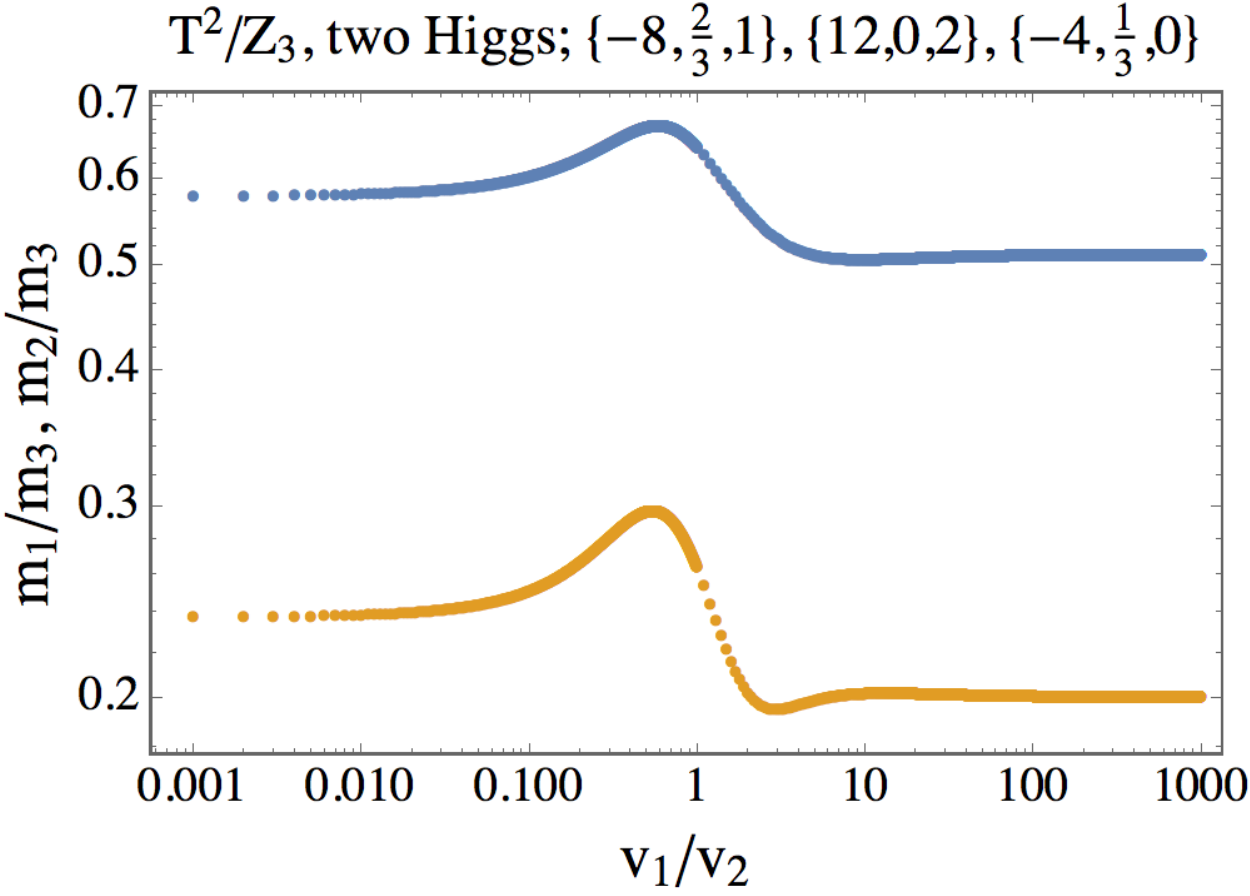} \\[6pt] 
\includegraphics[width=0.32\columnwidth]{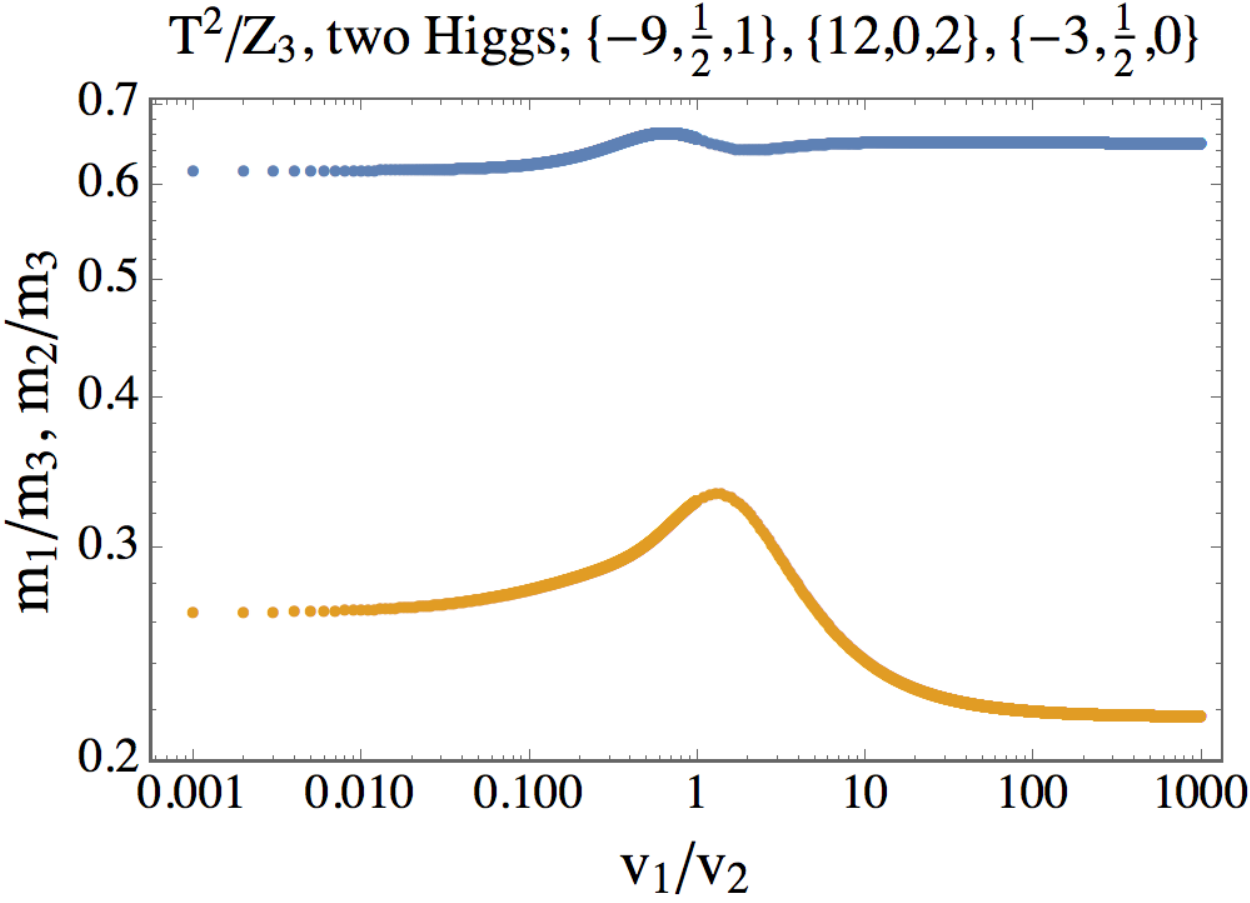}\
\includegraphics[width=0.32\columnwidth]{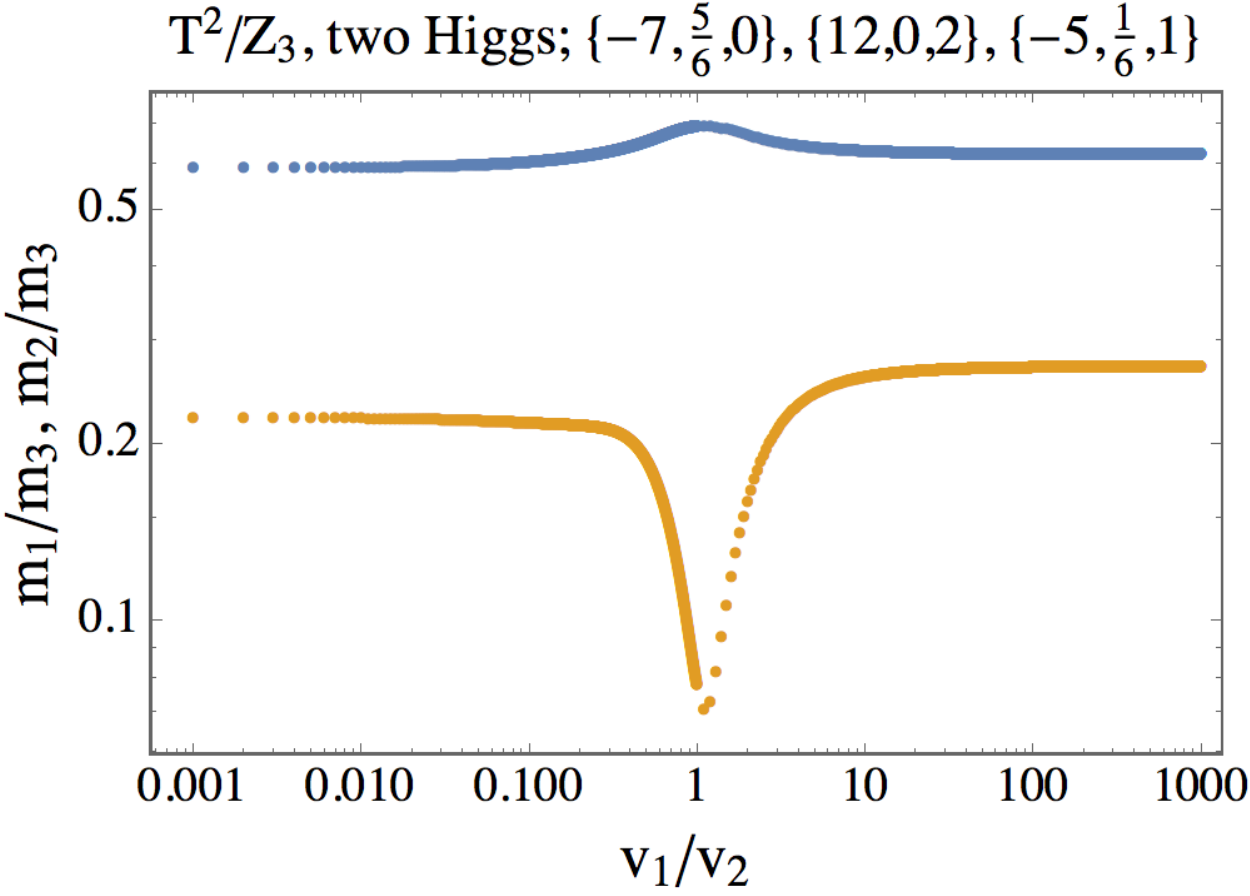}
\includegraphics[width=0.32\columnwidth]{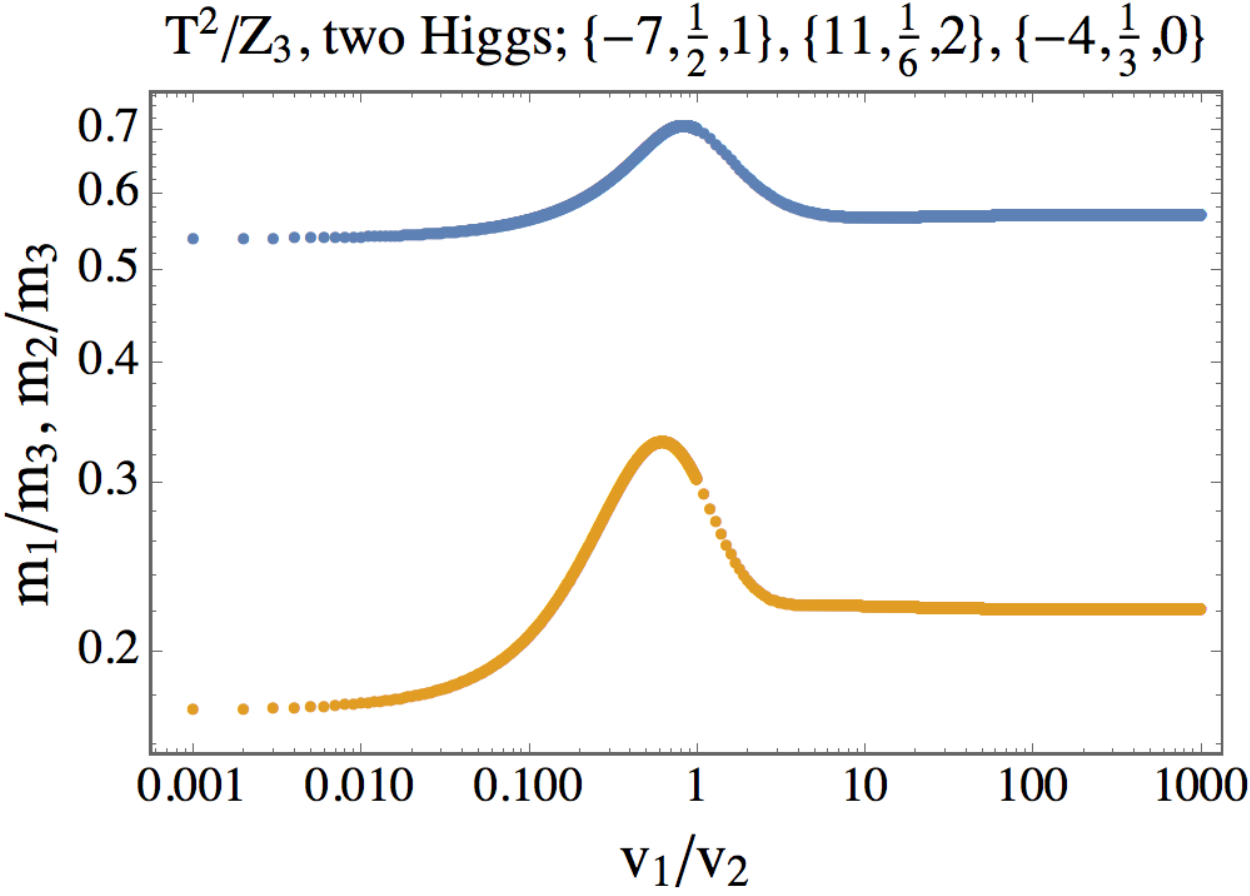}
\caption{
{Distributions of realized mass eigenvalues are shown when two Higgs bosons appear in the cases of $T^{2}/Z_{3}$.
We analyze all the cases (nine patterns in total) {where} rank-three mass matrices are realized.
Configurations are summarized as $\{ M_{ab}, \alpha_{ab}, s_{ab} \}$, $\{ M_{ca}, \alpha_{ca}, s_{ca} \}$, $\{ M_{bc}, \alpha_{bc}, s_{bc} \}${,} where we define $\eta_{ab} \equiv e^{2\pi i s_{ab}/3}$. The same holds for {the $bc$ and $ca$} sectors.
The orange (blue) dots correspond to the mass ratio $m_{1}/m_{3}$ ($m_{2}/m_{3}$) under the ordering $m_{1} \leq m_{2} \leq m_{3}$.}
}
\label{fig:result_twoHiggs_Z3}
\end{figure}

\begin{figure}[H]
\centering
\includegraphics[width=0.32\columnwidth]{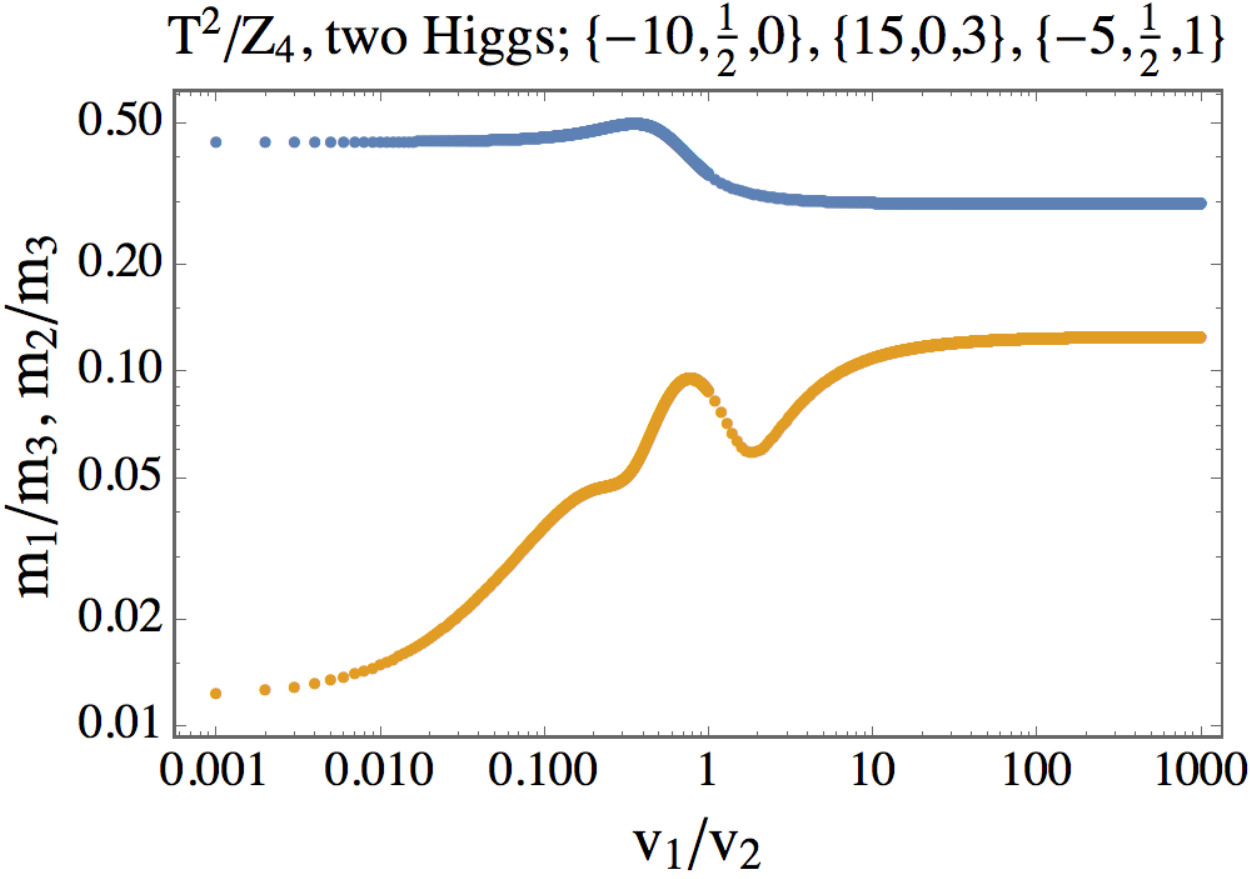}\ 
\includegraphics[width=0.32\columnwidth]{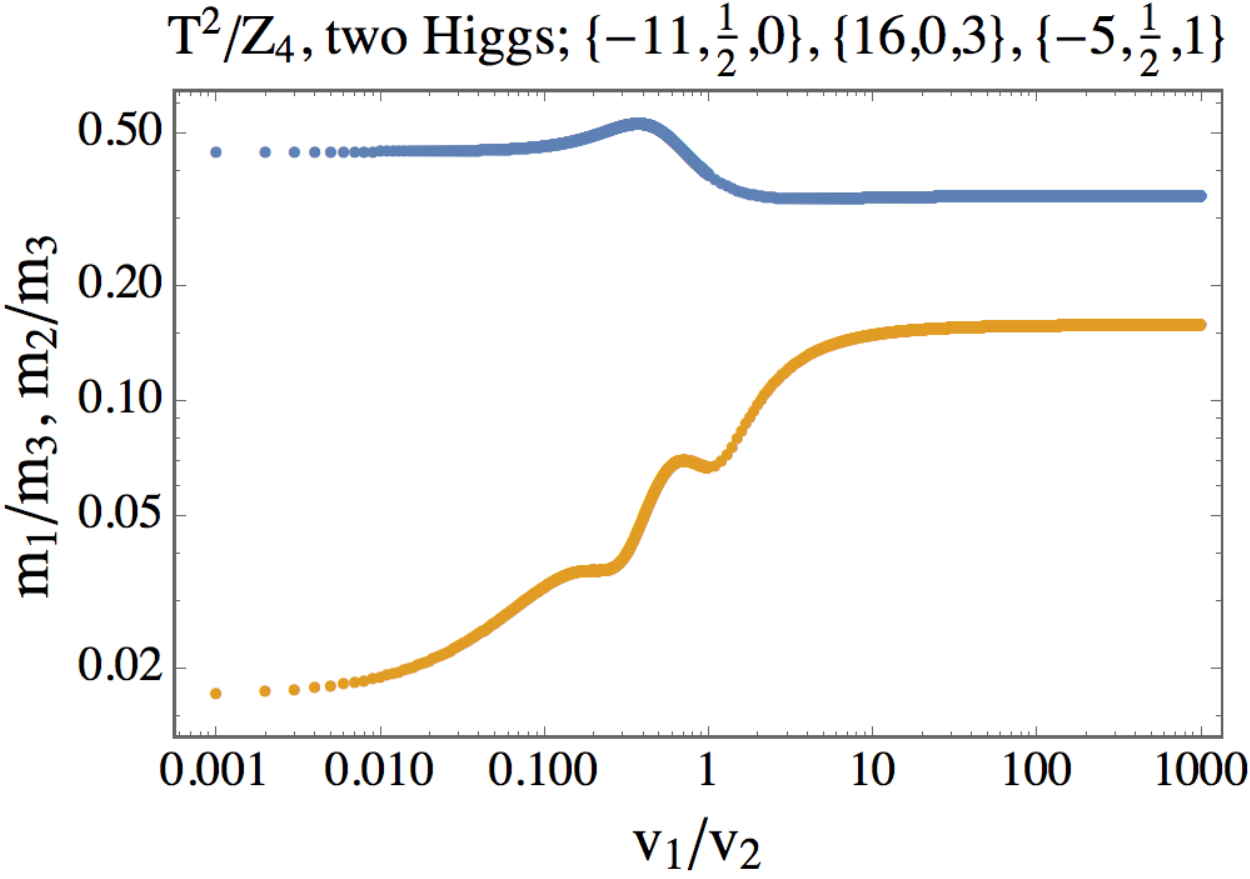}\
\includegraphics[width=0.32\columnwidth]{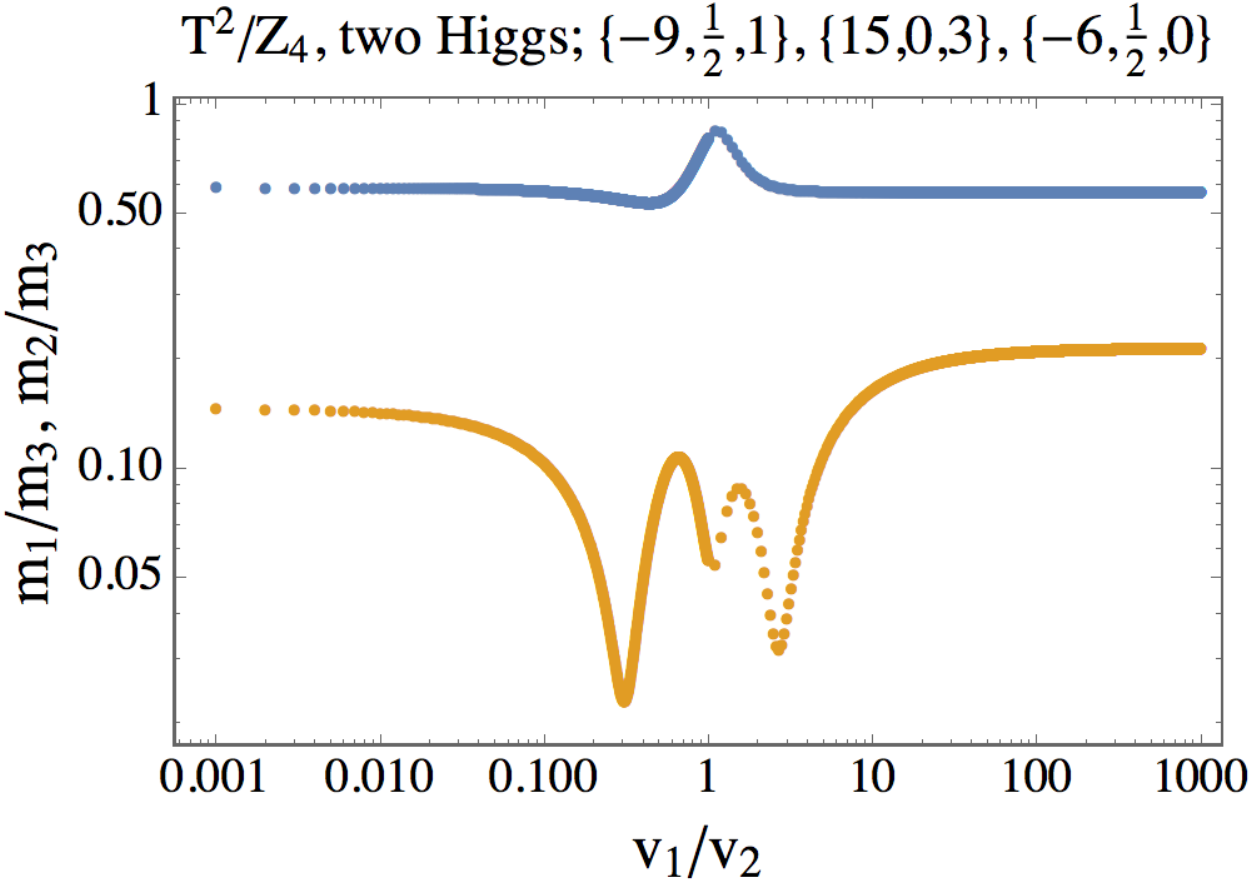} \\[6pt]
\includegraphics[width=0.32\columnwidth]{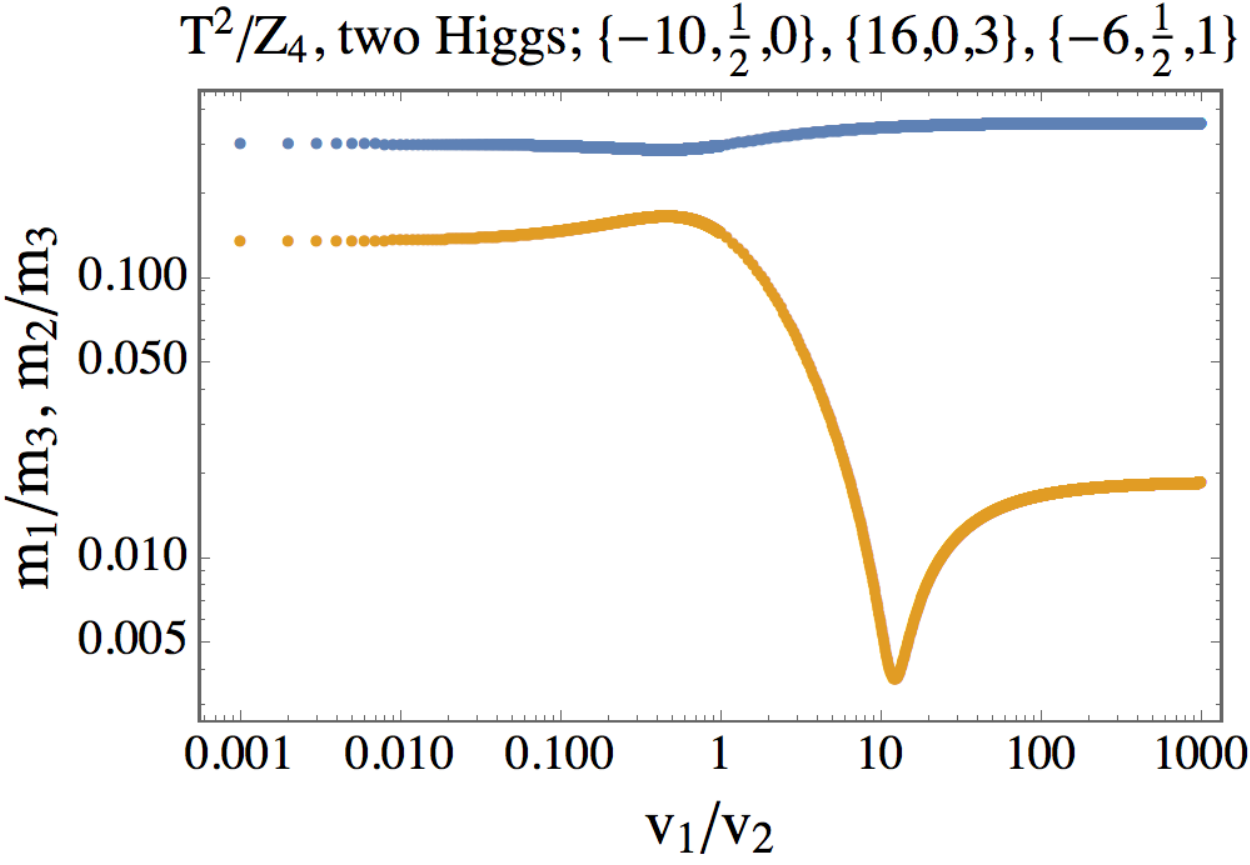}\ 
\includegraphics[width=0.32\columnwidth]{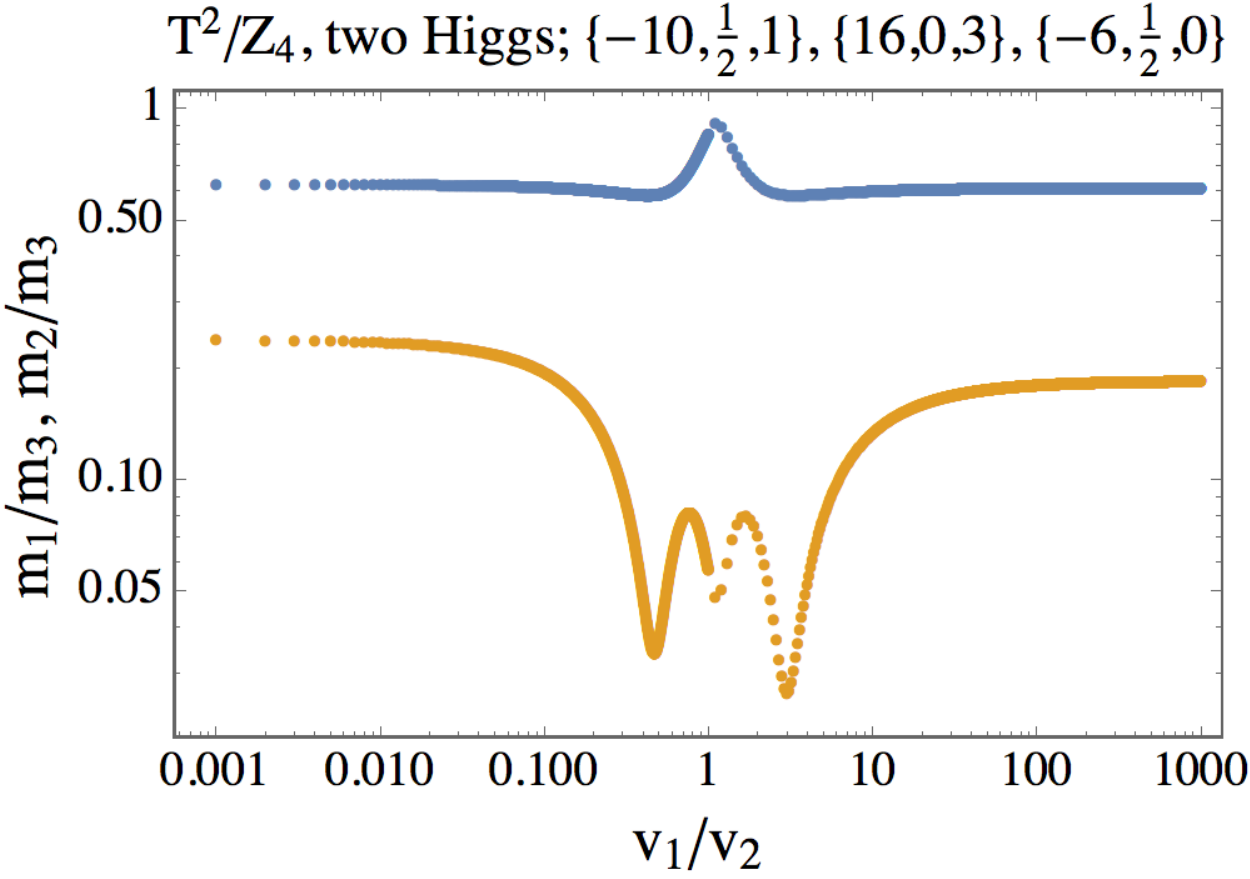}\
\includegraphics[width=0.32\columnwidth]{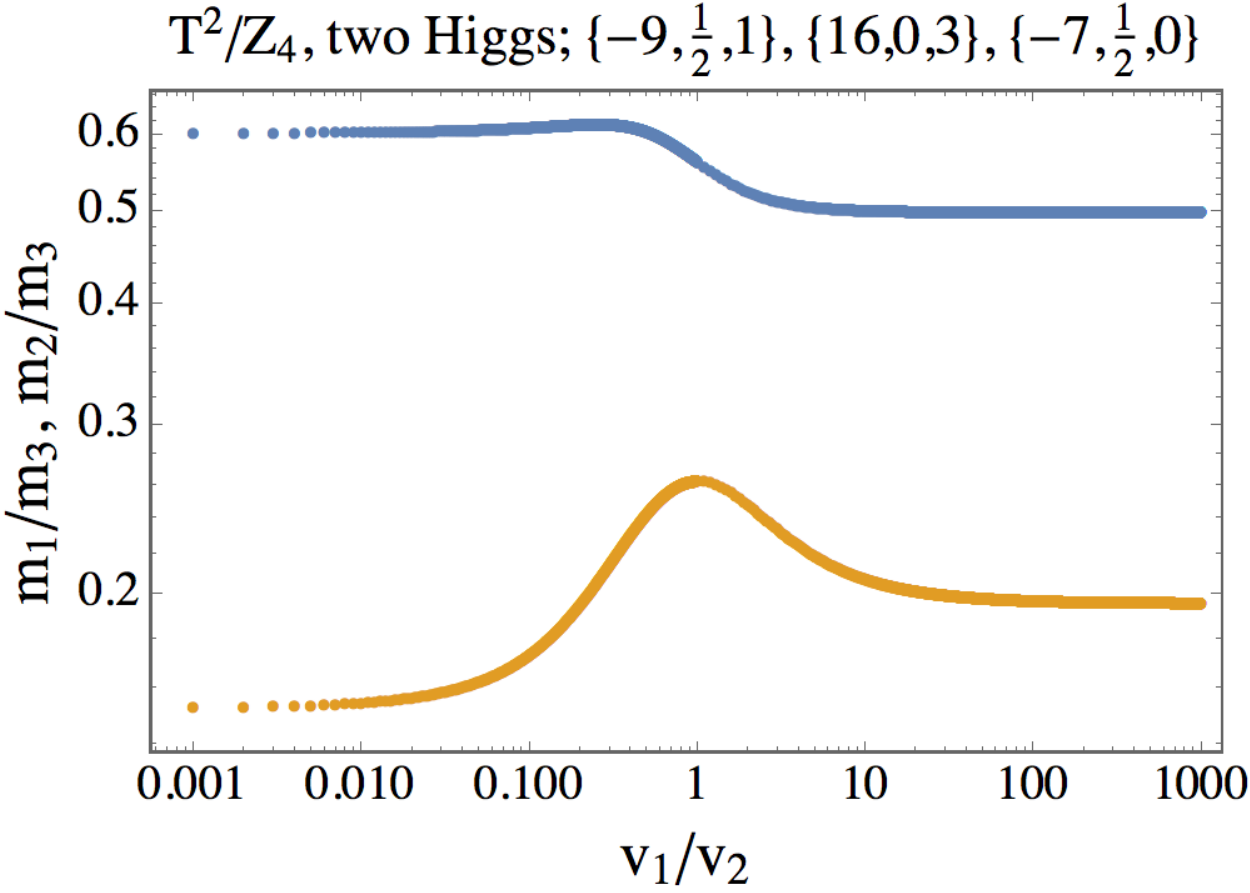}
\caption{
Distributions of realized mass eigenvalues are shown when two Higgs bosons appear in the cases of $T^{2}/Z_{4}$.
{We analyze all the cases (six patterns in total) {where} rank-three mass matrices are realized.}
Configurations are summarized as $\{ M_{ab}, \alpha_{ab}, s_{ab} \}$, $\{ M_{ca}, \alpha_{ca}, s_{ca} \}$, $\{ M_{bc}, \alpha_{bc}, s_{bc} \}${,} where we define $\eta_{ab} \equiv e^{2\pi i s_{ab}/4}$. The same holds for {the $bc$ and $ca$} sectors.
Conventions are the same as {those} adopted in Fig.~\ref{fig:result_twoHiggs_Z3}.
}
\label{fig:result_twoHiggs_Z4}
\end{figure}

\begin{figure}[H]
\centering
\includegraphics[width=0.32\columnwidth]{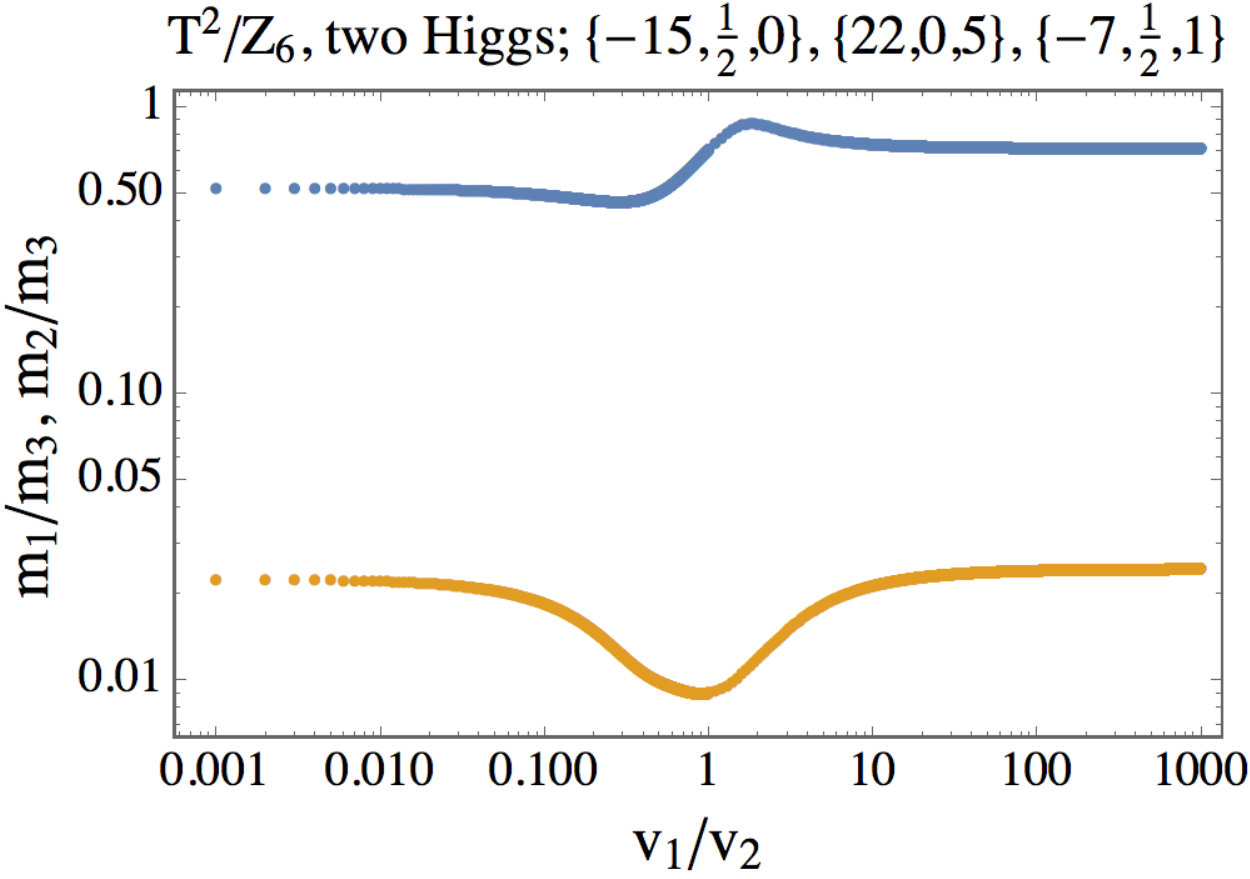}\ 
\includegraphics[width=0.32\columnwidth]{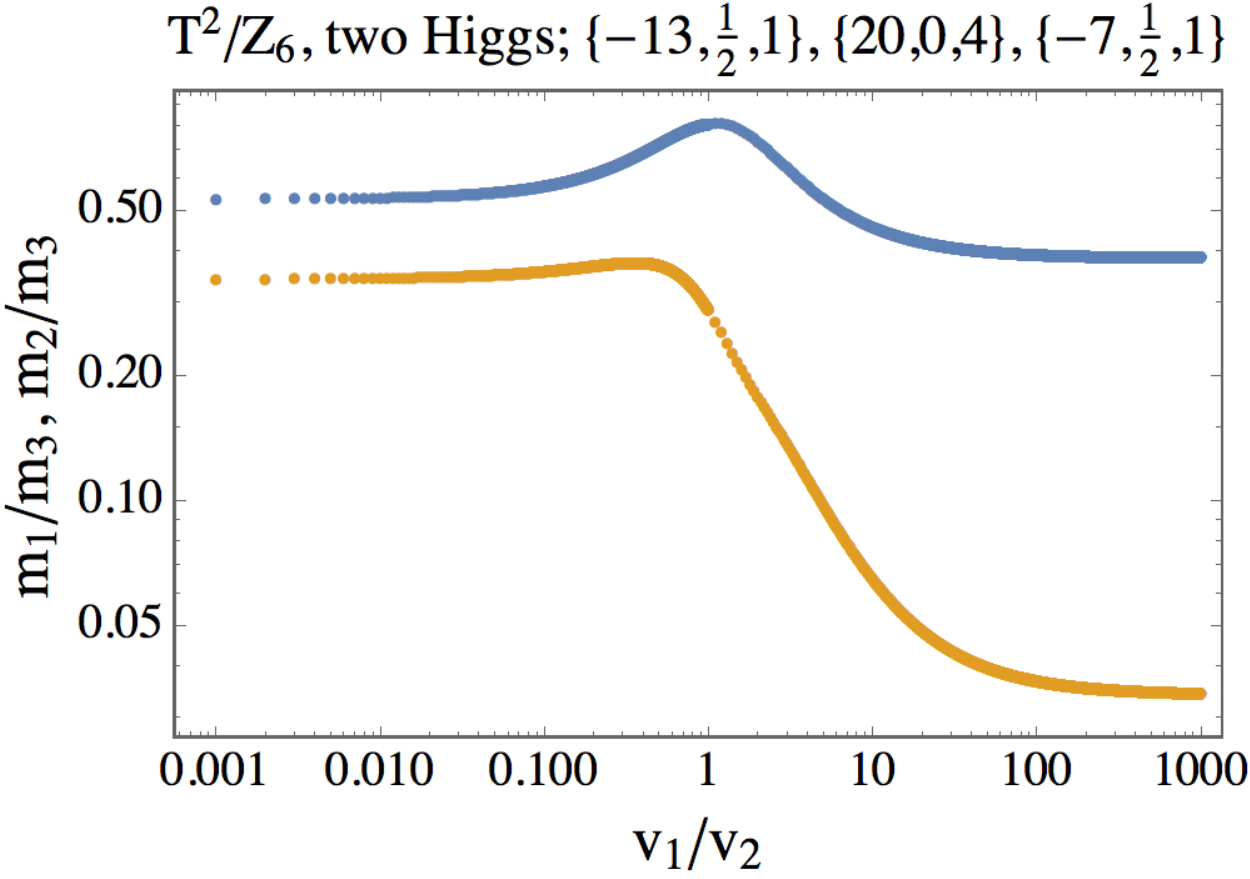}\
\includegraphics[width=0.32\columnwidth]{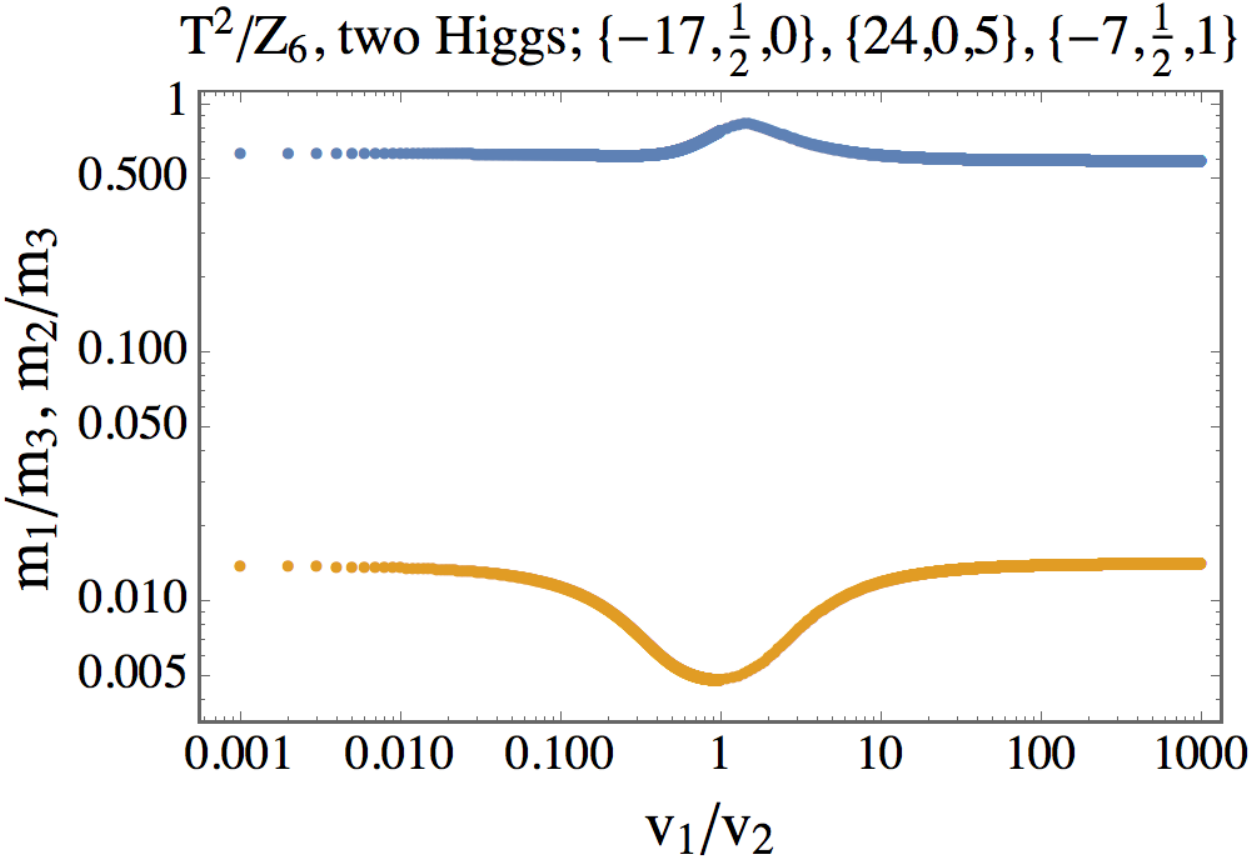} \\[6pt]
\includegraphics[width=0.32\columnwidth]{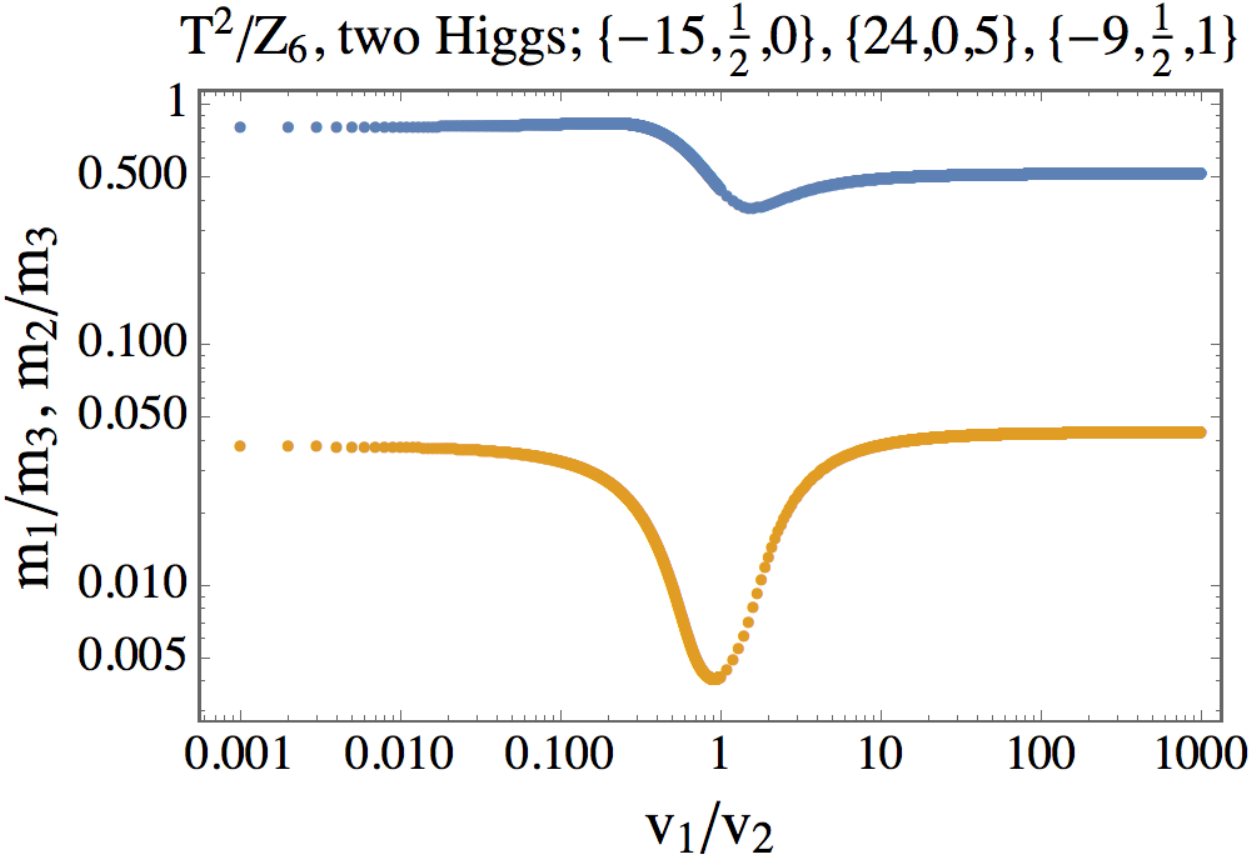}\ 
\includegraphics[width=0.32\columnwidth]{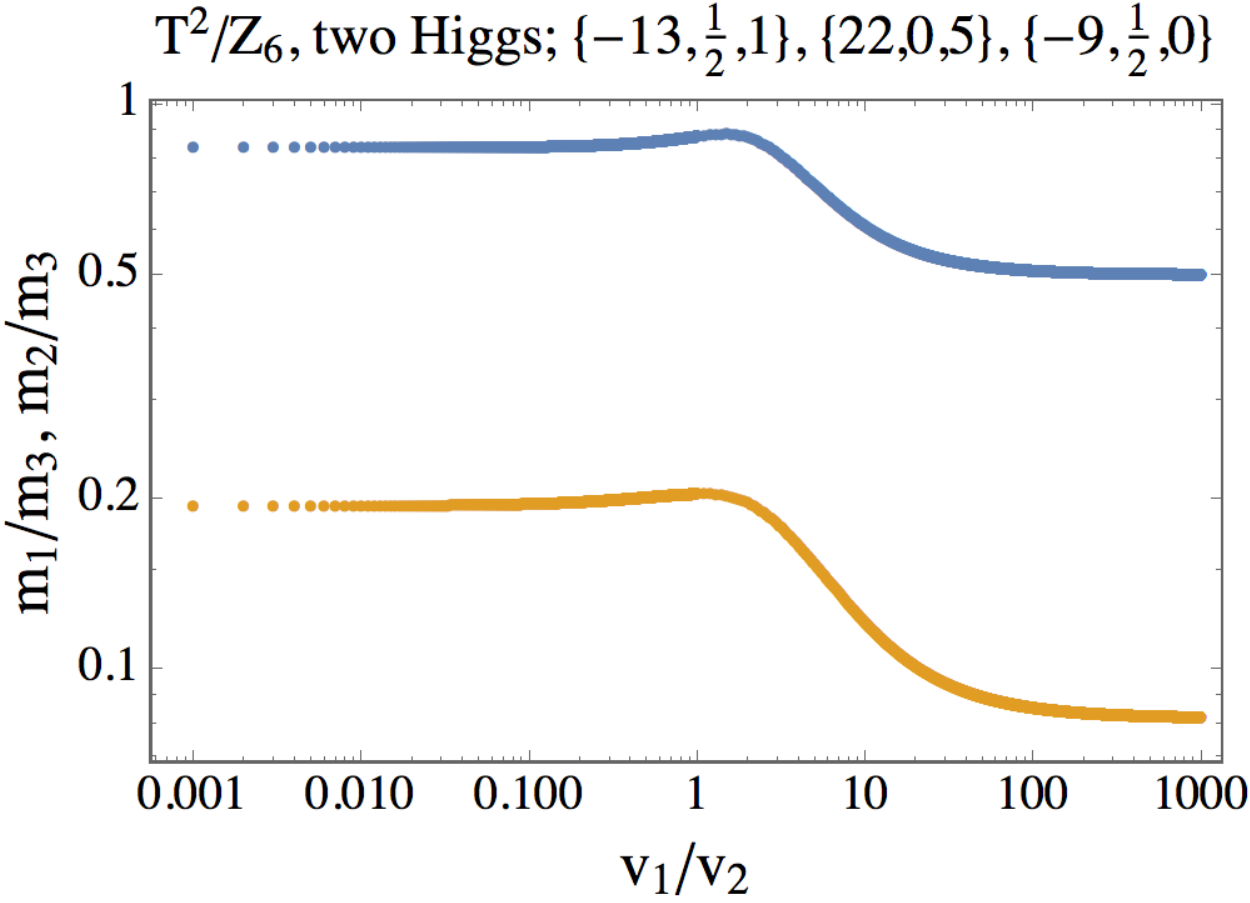} \\[6pt]
\includegraphics[width=0.32\columnwidth]{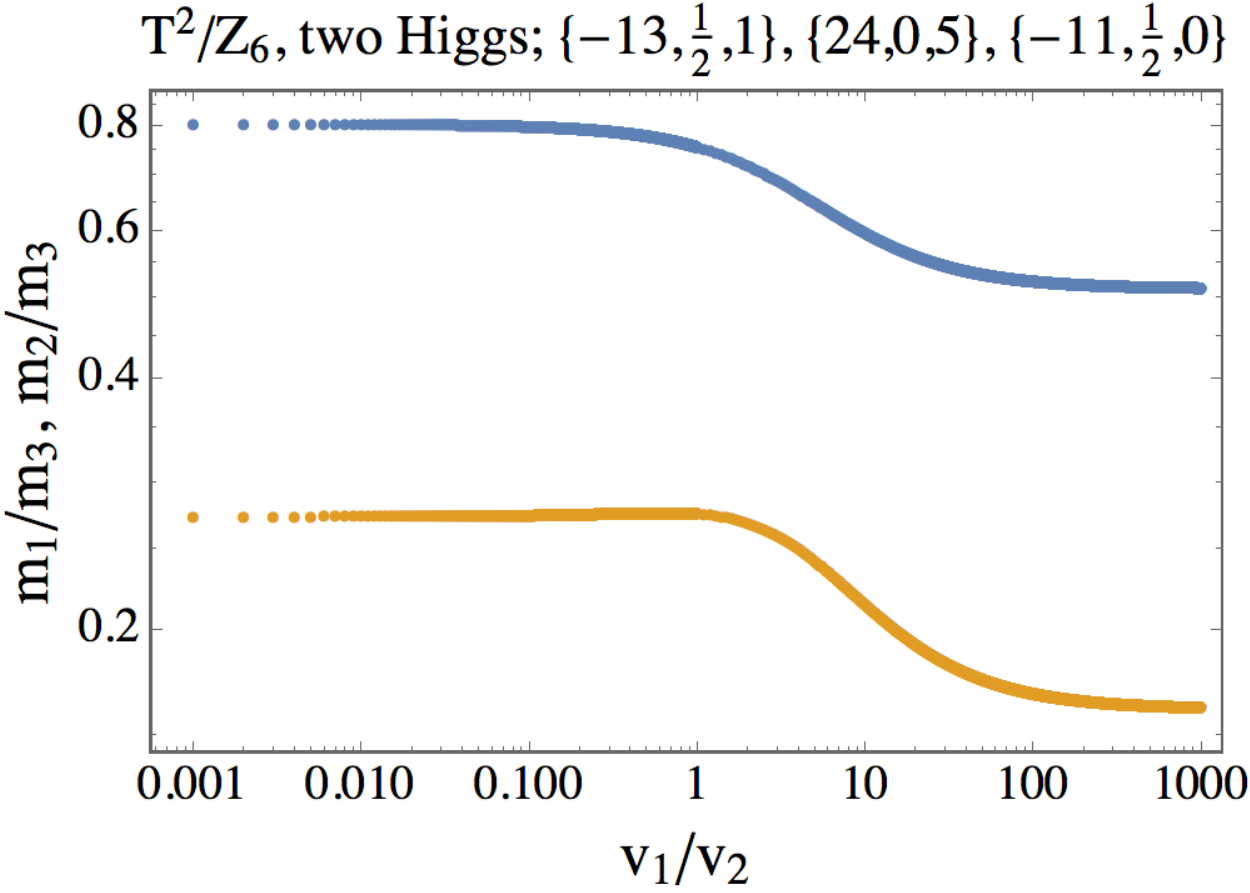}\  
\includegraphics[width=0.32\columnwidth]{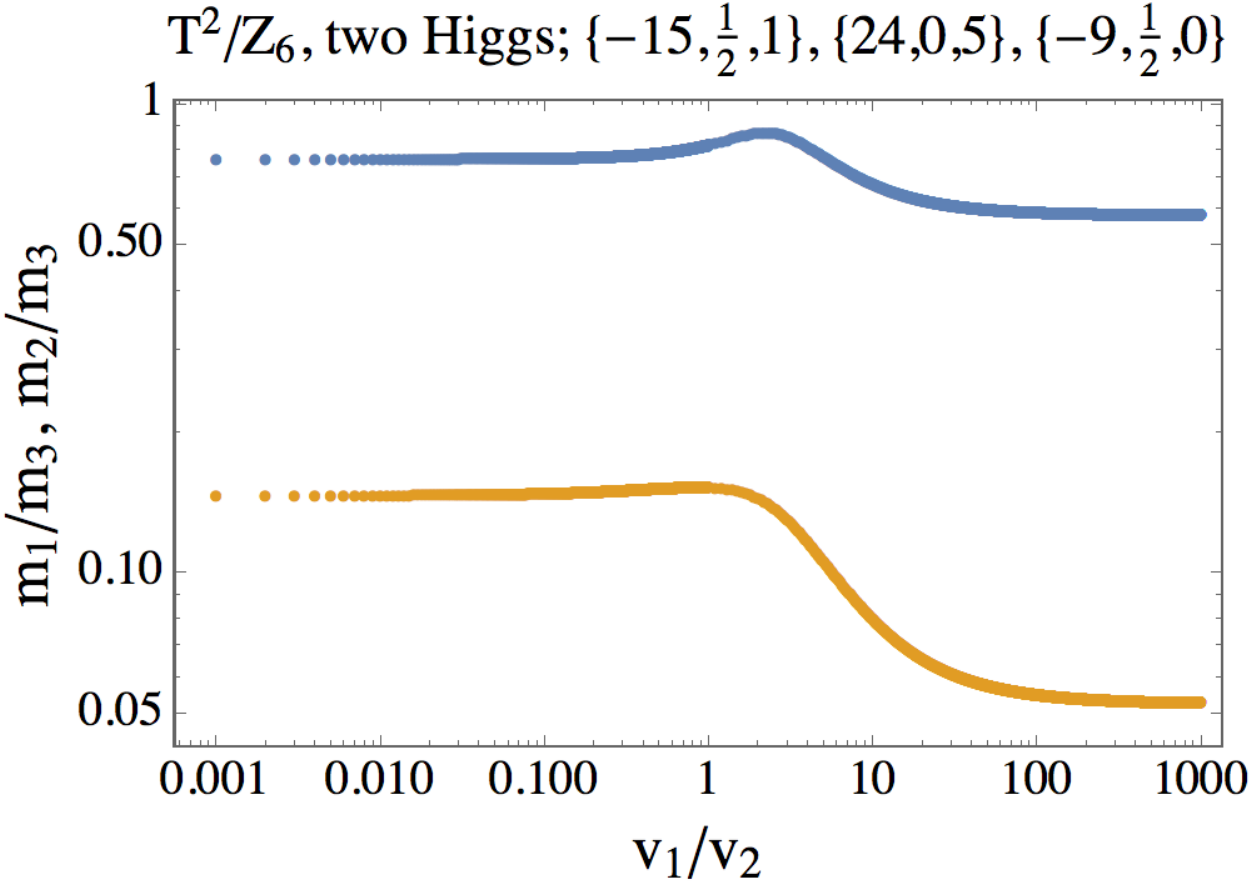}\
\caption{
Distributions of realized mass eigenvalues are shown when two Higgs bosons appear in the cases of $T^{2}/Z_{6}$.
{We analyze all the cases (seven patterns in total) {where} rank-three mass matrices are realized.}
Configurations are summarized as $\{ M_{ab}, \alpha_{ab}, s_{ab} \}$, $\{ M_{ca}, \alpha_{ca}, s_{ca} \}$, $\{ M_{bc}, \alpha_{bc}, s_{bc} \}${,} where we define $\eta_{ab} \equiv e^{2\pi i s_{ab}/6}$. The same holds for {the $bc$ and $ca$} sectors.
Conventions are the same as adopted {those} in Fig.~\ref{fig:result_twoHiggs_Z3}.
}
\label{fig:result_twoHiggs_Z6}
\end{figure}

The results are summarized in {Figs.}~\ref{fig:result_twoHiggs_Z3} $(T^{2}/Z_{3})$, \ref{fig:result_twoHiggs_Z4} $(T^{2}/Z_{4})$ and \ref{fig:result_twoHiggs_Z6} $(T^{2}/Z_{6})$, where intervals of dots (showing choices of $v_{1}/v_{2}$) are $0.001$ from $0.001$ to $1$, and $0.1$ from $1$ to $1000$, respectively.
Note that in the region where $v_{1}/v_{2}$ is more than $1000$ or less than $0.001$, either of {the} two mass matrices dominates and the two mass ratios are saturated.
Roughly speaking, the minimum values of $m_{1}/m_{3}$ among possible configurations are around $5 \times 10^{-2}$ for $T^{2}/Z_{3}$ and $5 \times 10^{-3}$ for $T^{2}/Z_{4,6}$, which are, of course, far from the required value $10^{-5}$.
Then, we conclude that it is hard to realize the ratio $m_{\text{up}}/m_{\text{top}}$ in every possibility with two Higgs pairs without {fine-tuning} in {$v_{1}/v_{2}$}.
Possibility would still remain when we accept {fine-tuning} in the ratio even through it is, at least to some extent, contradict to the basic motivation for considering such magnetized backgrounds in extra dimensions.\footnote{
In $N_{H} = 5$, the authors of~\cite{Abe:2014vza} found a configuration where all of the VEV ratios are within a natural range $[0.1,10]$.
}
{However,} if the mass hierarchy itself can become achievable under {fine-tuning}, we should justify the origin by dynamics.

We point out that the realized spectra are {nonlinearly} changed when we adjust the value of {$v_{1}/v_{2}$}.
Especially around $v_{1}/v_{2} = 1$, values of $m_{1}/m_{3}$ and $m_{2}/m_{3}$ alter significantly since elements of two mass matrices are comparable due to the mixing effect.
As an example, we show explicit forms of two mass {matrices} in $T^{2}/Z_{6}$ with quantum numbers $\{ M_{ab}, \alpha_{ab}, s_{ab} \}$, $\{ M_{ca}, \alpha_{ca}, s_{ca} \}$, $\{ M_{bc}, \alpha_{bc}, s_{bc} \}$ $=$ $\{-15,1/2,0\}, \{24,0,5\}, \{-9,1/2,1\}$:
\al{
\mathcal{M}_{1} &= v_{1}
\left(
\begin{array}{ccc}
 -0.204991-0.0796877 i & 0.00942303\, +0.09068 i & 0.0519518\, +0.0517703 i \\
 0.0763198\, +0.0222452 i & -0.00676905-0.017782 i & -0.0077252-0.00374813 i \\
 -0.0507492-0.0180038 i & 0.0683165\, +0.0433135 i & -0.0157981-0.12252 i \\
\end{array}
\right),
\label{eq:massmatrix_example_1} \\
\mathcal{M}_{2} &= v_{2}
\left(
\begin{array}{ccc}
 0.00505039\, +0.0102538 i & 0.00941751\, -0.180948 i & -0.0776918-0.115464 i \\
 -0.00982989-0.00704602 i & 0.00161259\, +0.0544036 i & 0.0166123\, +0.0174854 i \\
 0.0133678\, -0.0127981 i & 0.0383659\, -0.114777 i & -0.00590916+0.172257 i \\
\end{array}
\right).
\label{eq:massmatrix_example_2} 
}
Here, we define $Z_6$ {parity}, $\eta_{ab} \equiv e^{2\pi i s_{ab}/6}$. 
The same holds for {the $bc$ and $ca$} sectors.
Note that elements in the mass matrix (\ref{eq:massmatrix_example_1}) as well as (\ref{eq:massmatrix_example_2})
have no strong hierarchy.\footnote{
{For simplicity in the} calculation, we ignore the overall normalization factor $(\mathcal{N}_{M_{I}} \mathcal{N}_{M_{J}} / \mathcal{N}_{M_{K}})$ in Eq.~(\ref{general_Yukawa_formula}), which manifestly does not affect the mass ratios{.}
}
That is due to {the} effects of kinetic mixing, which {smears} hierarchies.
Obviously, the mass matrix (\ref{eq:massmatrix_example_1}) [(\ref{eq:massmatrix_example_2})] is dominant
when $v_1 \gg v_2$ [$v_1 \ll v_2$].
However, when $v_1 \sim v_2$,
partial cancellations happen sizably in {the} calculation of mass eigenvalues and the ratio $m_{1}/m_{3}$ drops {below} $0.005$ as shown in Fig.~\ref{fig:result_twoHiggs_Z6}.
The relatively hierarchical nature in $T^{2}/Z_{4,6}$ compared with $T^{2}/Z_{3}$ would originate from the relatively large magnitude of magnetic fluxes.

\section{Conclusions
\label{sec:conclusion}}

In this paper, we {discussed} how large mass hierarchy is realized on the magnetized extra dimension with orbifolding of $T^{2}/Z_{2}$, $T^{2}/Z_{3}$, $T^{2}/Z_{4}$ and $T^{2}/Z_{6}$.
We calculated realized mass eigenvalues in all the possibilities {that} predict three generations in zero mode fermions with one and two pairs of $SU(2)_{L}$ Higgs doublets (for up-type and down-type fermions).
{In $T^{2}/Z_{3,4,6}$, the effect of {the} kinetic mixings is nontrivial in {the} Yukawa calculation in Eq.~(\ref{eq:final_formula_Yukawa}) since it smears hierarchies in the mass matrices.
This feature brings us to the conclusion that it is very difficult to realize the mass ratio among the up quark and the top quark $m_{\text{up}}/m_{\text{top}} \sim 10^{-5}$ in all the {configurations} with one Higgs pair and with two Higgs pairs.
In the two Higgs cases, interference effects among two mass matrices are maximized at around $v_{1}/v_{2} = 1$ because elements of two matrices are comparable due to the mixing effect.
Situations would be the same when more Higgs doublets are realized on $T^{2}/Z_{3,4,6}$ where kinetic {mixing} smearing hierarchies would prevent a large mass difference like $m_{\text{up}}/m_{\text{top}} \sim 10^{-5}$.
Such obstacles are absent on $T^{2}$ and $T^{2}/Z_{2}$.
This information is very useful when we try to construct an actual model on magnetized extra dimensions.}


\section*{Acknowledgments}

This work is supported in part by {Grants-in-Aid} for Scientific Research [No.~25400252 and No.~26247042 (T.K.),  {No.~15K05055 and No.~25400260 (M.S.)}, {and} No.~16J04612 (Y.T.)] from the Ministry of Education, Culture, Sports, Science and Technology (MEXT) in Japan.


\bibliographystyle{utphys}
\bibliography{10D_comprehensive_Yukawa}

\end{document}